\documentclass[oneside]{ar2e}

\usepackage{graphicx} 
\usepackage{verbatim} 
\usepackage{bm}
\usepackage{color}

\begin{document}

\input epsf.tex    
\input epsf.def   

\input psfig.sty

\jname{Annu. Rev. Nucl. Part. Sci}
\jyear{2013}
\jvol{63}
\ARinfo{1056-8700/97/0610-00}

\def\snn{\mbox{$\sqrt{s_{_{\rm NN}}}$}}   
\def\pt{\mbox{$p_T$}}
\newcommand{\gapp}{\,{\raisebox{-.2ex}{$\stackrel{>}{_\sim}$}}\,}
\newcommand{\lapp}{\,{\raisebox{-.2ex}{$\stackrel{<}{_\sim}$}}\,}
\newcommand{\dla}{\langle\!\langle}
\newcommand{\dra}{\rangle\!\rangle}

\definecolor{dgreen}{cmyk}{1.,0.,1.,0.4}        
\definecolor{orange}{cmyk}{0.,0.353,1.,0.}    
\newcommand{\orange}[1]{\textcolor{orange}{#1}}
\newcommand{\blue}[1]{{\color{blue}{#1}}}
\newcommand{\green}[1]{{\color{green}{#1}}}
\newcommand{\red}[1]{{\color{red}{#1}}}
\newcommand{\magenta}[1]{{\color{magenta}{#1}}}
\def \new {\blue}
\def \old {\orange}
\def \ask {\magenta}


\title{Collective Flow and Viscosity in Relativistic Heavy-Ion Collisions}

\markboth{Heinz \& Snellings}{Flow and Viscosity in Relativistic Heavy Ion Collisions}

\author{Ulrich Heinz$^1$ and Raimond Snellings$^2$
\affiliation{$^1$Physics Department, The Ohio State University, Columbus, OH 43210-1117, USA;
email: heinz@mps.ohio-state.edu\\
$^2$Physics Department, Utrecht University, NL-3584 CC Utrecht, The Netherlands; email:
R.J.M.Snellings@uu.nl}
}


\begin{abstract}
Collective flow, its anisotropies and its event-to-event fluctuations in relativistic heavy-ion collisions, and the extraction of the specific shear viscosity of quark-gluon plasma (QGP) from collective flow data collected in heavy-ion collision experiments at RHIC and LHC are reviewed.
Specific emphasis is placed on the similarities between the Big Bang of our universe and the Little Bangs created in heavy-ion collisions. 
\end{abstract}

\maketitle

\noindent

\section{HISTORICAL OVERVIEW}
\label{overview}

The idea that high energy collisions between hadrons and atomic nuclei can be described macroscopically using relativistic fluid dynamics has been around for over half a century \cite{Fermi:1950jd}. It gained strength in the late 1970's to mid 1990's as fixed-target heavy-ion collision experiments at beam energies from a few hundred MeV to 200 GeV per nucleon revealed unmistakable evidence for the formation of dense matter that underwent collective expansion in the directions perpendicular to the beam (``transverse flow'') \cite{Nagamiya:1981sd}. The discovery of transverse, in particular {\em elliptic} flow (anisotropic emission around the beam direction) \cite{Gustafsson:1984ka,Appelshauser:1997dg} led to the theoretical development of numerical codes that solved the equations of relativistic ideal fluid dynamics in one and two transverse directions, using coordinates and initial conditions appropriate for relativistic heavy-ion collisions \cite{Schlei:1995hv}. The predictions from such models for the momentum distributions of the finally emitted hadrons reproduced qualitatively all experimentally observed features of soft (i.e. low transverse momentum) particle production in heavy-ion collisions \cite{Kolb:2003dz}, but failed in important aspects quantitatively (for example, hydrodynamics overpredicted the observed elliptic flow by about 50\% \cite{Kolb:2000fha} at CERN SPS energies while microscopic approaches based on a kinetic description of systems of scattering hadrons got it about right \cite{Soff:1999yg} but seriously underpredicted it later at higher energies \cite{Bleicher:2000sx}).

In 2000 the Relativistic Heavy Ion Collider (RHIC) began operation at ten times larger center of mass energy than previously available. It was built to mass-produce quark-gluon plasma (QGP), a new type of dense matter in which hadrons dissolve into deconfined colored degrees of freedom, quarks and gluons, that was predicted by the modern theory of the strong interaction, Quantum Chromodynamics (QCD). RHIC data appeared to agree, for the first time, {\em quantitatively} with predictions from ideal fluid dynamics \cite{Kolb:2003dz}. This caused a paradigm change in the field: Having expected gas-like behavior, based on the ideas of asymptotic freedom and color Debye screening in the QGP, the community was forced to accept instead the notion that the QGP is a strongly-coupled plasma that flows like a liquid \cite{Heinz:2001xi,Gyulassy:2004vg}.

Later it became clear that some of the initial success of ideal fluid dynamics was artificial and due to the use of an inadequate equation of state for the fireball matter \cite{Huovinen:2005gy} as well as an incorrect treatment of the chemical composition of the fireball during its late hadronic stage \cite{Hirano:2002ds}. A key insight was \cite{Bass:2000ib,Teaney:2001av,Hirano:2005xf} that the hydrodynamic description should be restricted to the dense quark-gluon plasma (QGP) stage of the collision whereas the late hadronic stage (after the recombination of colored quarks and gluons into color-neutral hadrons) is too dissipative for a fluid dynamical approach \cite{Song:2010aq} and must be described microscopically. Hybrid codes that couple (ideal) fluid dynamics of the QGP to a microscopic hadron cascade worked better than a purely hydrodynamic approach \cite{Bass:2000ib,Teaney:2001av,Hirano:2005xf} and explained why pure fluid dynamics was quantitatively successful only in central collisions between large ($A{\,\sim\,}200$) nuclei at midrapidity at top RHIC energy, but gradually broke down in smaller collision systems, in more peripheral collisions, away from midrapidity, and at lower collision energies \cite{Heinz:2004ar}. 

This observation focused the attention of the community on the importance of dissipative effects which had thus far been ignored. With improving experimental precision it became increasingly obvious that even in the most central collisions among the heaviest available nuclei the data required some degree of QGP viscosity \cite{Hirano:2005xf,Lacey:2006pn,Romatschke:2007mq}. The basic success of the ideal fluid approach suggested, however, that this viscosity had to be small. This begged the question ``How small exactly?" which marked the transition of the heavy-ion program from its initial QGP discovery phase to a second stage of quantitative characterization of the QGP.

Theoretical work on strongly coupled quantum field theories, borrowing tools from superstring theory, established a lower limit around $1/4\pi$ for the specific shear viscosity $\eta/s$ of the QGP \cite{Policastro:2001yc} (see Chpt.~\ref{hydro}). How to extract this quantity from experimental data will be described in this review. In this extraction a key role is played by the fact that the initial conditions in heavy-ion collisions fluctuate from event to event \cite{Miller:2003kd,Socolowski:2004hw,Sorensen:2010zq,Alver:2010gr}. Just as in the Big Bang gravity evolves the initial density fluctuations imprinted on the cosmic microwave background \cite{Komatsu:2010fb} into today's star and galaxy distributions \cite{Springel:2006vs}, viscous hydrodynamics converts the initial energy density fluctuations of the ``Little Bangs'' created in heavy-ion collisions into a distribution of anisotropic harmonic flow components \cite{Mishra:2007tw,Sorensen:2010zq,Mocsy:2011xx,Qin:2010pf,Qiu:2011iv}. The low shear viscosity of the QGP allows traces of the initial fluctuations to survive to the final freeze-out stage, i.e. the flow anisotropies generated by them are not completely erased by dissipation and can be used to determine the QGP viscosity. While we have observational access only to a single instance of the Big Bang (the one that created the universe in which we live), RHIC and the Large Hadron Collider (LHC) have created billions of Little Bangs, each with a different distribution of initial fluctuations and its own hydrodynamic response to it. From this extraordinarily rich data set both the QGP transport coefficients and the initial fluctuation spectrum can be reconstructed with precision. While this task is not yet complete, we review the dramatic progress and significant success achieved over the last few years.

\section{RELATIVISTIC HEAVY-ION COLLISION DYNAMICS}
\label{hydro}
\subsection{Second order viscous relativistic fluid dynamics}
\label{sec2.1}

Hydrodynamics is an effective macroscopic description of a system that is in approximate local thermal equilibrium. It can be derived from the underlying microscopic (kinetic) description through an expansion in gradients of the local thermodynamic variables. In zeroth order of gradients one obtains ideal fluid dynamics. Navier-Stokes theory accounts for terms that are linear in gradients. These describe momentum and baryon number transport across fluid cells due to bulk and shear viscosity as well as heat conduction. Relativistic Navier-Stokes theory is, unfortunately, acausal \cite{Hiscock:1983zz}: superluminal signal propagation causes short wavelength perturbations (which are outside the validity range of the gradient expansion) to grow exponentially, leading to physical and numerical instabilities. Any numerical implementation of relativistic dissipative fluid dynamics thus requires the inclusion of terms that are second order in gradients. Some of the associated transport coefficients have the physical meaning of microscopic relaxation times for the dissipative flows. Like the viscosities and the heat conductivity, they depend on the microscopic scattering dynamics; they encode the time delay between the appearance of thermodynamic gradients that drive the system out of local equilibrium and the associated build-up of dissipative flows in response to these gradients, thereby restoring causality. 

The most widely used theoretical frameworks for second-order dissipative relativistic fluid dynamics are the 35-year-old Israel-Stewart theory \cite{Israel:1976tn} and its recent generalizations \cite{Baier:2007ix,Muronga:2003ta,Heinz:2005bw,Song:2007fn,Song:2008si,Molnar:2009tx,Denicol:2010xn}. In its full glory, this theory contains a large number of second-order terms \cite{Denicol:2012cn}, but none of the present numerical implementations includes all of them. [Some exploratory studies including third-order gradient terms have also been published \cite{El:2009vj}, and they appear to improve the convergence of the gradient expansion.] While a full investigation is still outstanding, spot checks indicate that in practice not all second-order terms are equally important. Accounting for non-zero relaxation times at all stages of the evolution is, however, crucial: they limit excursions from local equilibrium, thereby both stabilizing the theory and improving its quantitative precision \cite{Song:2007fn,Song:2008si}. 

For a system with small or zero net baryon number, heat conduction effects can be neglected. This simplification is generally being employed at LHC and top RHIC energies but will have to be abandoned at lower energies. This leaves the bulk and shear viscosities, $\zeta$ and $\eta$, and the relaxation times $\tau_\zeta$ and $\tau_\eta$ for the bulk and shear viscous pressures as the main transport coefficients controlling the collective dynamics. They can be usefully characterized by their unitless combinations with the entropy density $s$ and the temperature $T$, resulting in the specific viscosities $\eta/s$ and $\zeta/s$ and the scaled relaxation times $T\tau_{\zeta,\eta}$. For consistency of the theory the microscopic relaxation rates must be much larger than the scalar expansion rate $\theta=\partial_\mu u^\mu$ of the fluid, $\tau_{\eta,\zeta}\theta{\,\ll\,}1$. ($u^\mu(x)$ is the flow 4-velocity profile of the fluid.)

\subsection{Equation of state (EOS)}
\label{sec2.2}

Following recent progress in lattice gauge theory, the model equations of state of the past have been increasingly replaced by parametrizations of the lattice QCD EOS for which now nicely converged results are available (at least for baryon-free systems) \cite{Borsanyi:2010cj}. Below $T_c{\,\simeq\,}155{-}160$\,MeV \cite{Borsanyi:2010cj} (the pseudo-critical temperature at which quarks and gluons combine into hadrons in a smooth but rapid crossover transition \cite{Aoki:2006we}) the lattice QCD EOS is matched to that of a hadron resonance gas \cite{Huovinen:2009yb}. What matters for the development of collective flow is the relationship between the pressure $p$ (whose gradients provide the accelerating forces) and the enthalpy $e{+}p$ which embodies the inertia of the fluid ($e$ is the comoving energy density). For ideal fluids (ignoring the viscous terms) the relativistic Euler equation can be written in the form $\dot{u}^\nu=\frac{c_2^2}{1{+}c_s^2}\frac{\nabla^\nu p}{p}$, where we expressed the EOS $p(e)$ approximately in terms of the squared speed of sound $c_s^2{\,=\,}\partial p/\partial e$ as $p{\,=\,}c_s^2 e$. This shows that the evolution of flow in response to pressure gradients is controlled by the stiffness $\partial p/\partial e$ of the EOS, i.e. the key hydrodynamic ingredient is the speed of sound as a function of local energy density, $c_s^2(e)$. The connection between energy density $e$ and temperature $T$ (which depends on the number of active degrees of freedom in the medium and thus on its microscopic composition) is not needed for the hydrodynamic evolution; it is required, however, for the calculation of thermal electromagnetic radiation from the expanding fireball and final particle emission whose emission rates and energy distributions depends on the local temperature $T$.

\subsection{Quark-gluon plasma transport coefficients}
\label{sec2.3}

The calculation of QGP transport coefficients from QCD is difficult. Very little is known from first principles about heat conductivity in QCD matter. Leading order results in a weak-coupling expansion ($g{\ll}1$) for the shear and bulk viscosities of a quark-gluon plasma were obtained in \cite{Arnold:2000dr,Arnold:2006fz}. For the coupling strengths relevant in a QGP at temperatures of a few $T_c$, $g\sim1-2$, these calculations provide not much more than an order of magnitude estimate. Taken literally, the leading-order perturbative value of the QGP shear viscosity \cite{Arnold:2000dr} is too large by a factor 3-10 to explain the experimentally measured large anisotropic flow coefficients and the correspondingly required short QGP thermalization times \cite{Csernai:2006zz,Kapusta:2008vb}. The extraction of $\eta$ and $\zeta$ directly from lattice QCD requires an analytic continuation from imaginary to real times. This is numerically costly and so far has only yielded results with large error bands \cite{Meyer:2007ic}.

An alternate approach exploits the AdS/CFT correspondence relating strong\-ly coupled conformal field theories (CFT) to classical gravity in weakly curved Anti-de-Sitter (AdS) space-time geometries \cite{Maldacena:1997re}. This approach has established approximate strong-coupling limits for the specific shear and bulk viscosities $\eta/s{\,\simeq\,}1/(4\pi)$ \cite{Policastro:2001yc} and $\zeta/\eta{\,\simeq\,}2\left(\frac{1}{3}{-}c_s^2\right)$ \cite{Buchel:2007mf}. For $\eta/s$ this is significantly below the perturbative QCD estimate while $\zeta/s$ (which is proportional to the interaction-induced deviation from the conformal limit, $\frac{1}{3}{-}c_s^2$) is much larger than in perturbative QCD where $\zeta/\eta\approx15\left(\frac{1}{3}{-}c_s^2\right)^2\sim g^8$ \cite{Arnold:2006fz}.

Below $T_c$, a number of calculations of $\eta/s$ and $\zeta/s$ based on effective hadronic interaction models (see Kapusta \cite{Kapusta:2008vb} for a review) and on transport models
\cite{Demir:2008tr} have been performed. Combining all calculations one finds that, generically, $\eta/s$ has a minimum near $T_c$, rising steeply below and slowly above $T_c$. The specific bulk viscosity $\zeta/s$, on the other hand, peaks near $T_c$ \cite{Karsch:2007jc}, due to the breaking of conformal symmetry by long-range critical correlations; above and below $T_c$, the ratio $\zeta/\eta$ generically decreases, allowing shear viscous effects to dominate over bulk viscous ones. The strongest effects from bulk viscosity (if any) are expected near $T_c$.

In view of limited theoretical knowledge of the QGP transport coefficients, recent attention has focussed on extracting them phenomenologically from experimental data, by comparison with viscous fluid dynamical simulations. 

\subsection{Hadron cascade stage, and chemical and kinetic freeze-out}
\label{sec2.4}

Below $T_c$ chemical reactions between different hadronic species become too slow to 
maintain chemical equilibrium in the exploding fireball \cite{Heinz:1999kb,BraunMunzinger:2003zz}. This leads to ``chemical freeze-out'': While maintaining some degree of local kinetic equilibrium through quasi-elastic resonance scattering, the final stable hadron yields hardly change any more. An important exception are the baryon and antibaryon abundances which are somewhat depleted by annihilation in the hadronic rescattering stage \cite{Bass:1999tu,Becattini:2012sq}. After correcting for baryon-antbaryon annihilation, the finally observed stable hadron yields reflect approximate chemical equilibrium abundances with a temperature $T_\mathrm{chem}$ close to $T_c$ \cite{Becattini:2012sq}. 

If one continues with viscous hydrodynamics to describe the hadronic phase below $T_c$, down to a final kinetic decoupling temperature $T_\mathrm{kin}<T_\mathrm{chem}$, one must account for chemical freeze-out at $T_\mathrm{chem}$ by assigning temperature-dependent non-equilibrium chemical potentials to each hadron species \cite{Hirano:2002ds,Bebie:1991ij}. It was found that an incorrect chemical composition at $T_\mathrm{kin}$ can seriously distort the distribution of the hydrodynamically generated momentum anisotropy over the different hadron species and over their transverse momenta \cite{Hirano:2002ds}. Without accounting for the non-equilibrium hadronic chemical composition below $T_\mathrm{chem}$, phenomenologically extracted values for the QGP shear viscosity can be wrong by 100\% \cite{Song:2008hj,Heinz:2009xj}.  

In spite of large resonant hadronic scattering cross sections, however, the collective expansion is so fast that even kinetic equilibrium is difficult to maintain below $T_c$. Once in the hadronic phase, the fluid becomes so dissipative \cite{Hirano:2005xf} that the macroscopic hydrodynamic description becomes unreliable \cite{Song:2010aq} and should be replaced by a microscopic solution of the coupled Boltzmann equations for the various hadronic phase-space distributions (``hadron cascade", see e.g. \cite{Bass:1998ca,Nara:1999dz}). Still, many studies today continue to use the fluid dynamical short-cut all the way down to $T_\mathrm{kin}$. In such an approach, final hadron distributions are computed by converting the hydrodynamic output on a hypersurface of constant decoupling temperature $T_\mathrm{kin}$ into hadron momentum distributions using the Cooper-Frye prescription, with non-equilibrium chemical potentials for the various hadron species and non-equilibrium corrections $\delta f(x,p)$ to the local phase-space distribution \cite{Israel:1976tn,Teaney:2003kp} that account for the viscous corrections to the energy momentum tensor on this decoupling surface. 

The more reliable, but much more expensive hybrid approach couples a viscous fluid dynamic description of the QGP phase with a microscopic Boltzmann simulation of the hadronic phase \cite{Song:2010aq,Ryu:2012at}. {\tt VISHNU} \cite{Song:2010aq} matches the (2+1)-dimensional, longitudinally boost-invariant viscous hydrodynamic algorithm {\tt VISH2{+1}} \cite{Song:2007fn,Song:2008si} to the well-known {\tt UrQMD} cascade \cite{Bass:1998ca},  in {\tt MUSIC}{+}{\tt UrQMD} \cite{Ryu:2012at} the evolution is fully (3+1)-dimensional. In these hybrid approaches the Cooper-Frye algorithm is used to convert hydrodynamic output into particle phase-space distributions on a switching surface of constant temperature $T_\mathrm{chem}$, assuming chemical equilibrium yields \cite{Song:2010aq}. {\tt UrQMD} then propagates these particles until all interactions cease and all unstable resonances have decayed. For good statistics, the final hadronic cascade stage is simulated many times for each hydrodynamic event.    

The influence of the dissipative hadronic stage on final observables increases relative to that of the early QGP stage as the collision energy decreases and the fireball spends less time as a QGP and a larger fraction of its history in the hadronic stage. Hybrid approaches such as {\tt VISHNU} thus become more and more important as one moves down in energy. Some features measured at lower energies during the recent RHIC beam energy scan, such as different differential elliptic flow coefficients $v_2(p_T)$ for baryons and antibaryons \cite{Dong:2012mt}, are difficult to understand in pure hydrodynamics, and a hybrid approach may prove essential for their interpretation.  

At the highest available collision energies at the LHC, most of the anisotropic flow is created before hadronization, and the relative impact of the late hadronic stage is weaker \cite{Hirano:2007xd}. Good descriptions of the charged hadron $p_T$-distributions and anisotropic flows $v_n$ and $v_n(p_T)$ have been obtained both with pure viscous hydrodynamics \cite{Shen:2011eg,Schenke:2011zz,Bozek:2012qs,Gale:2012rq} and with the hybrid code {\tt VISHNU} \cite{Heinz:2011kt} (see also recent work with {\tt MUSIC}{+}{\tt UrQMD} \cite{Ryu:2012at} and with hybrid codes coupling ideal fluid dynamics with {\tt UrQMD} \cite{Hirano:2010je,Akkelin:2008eh}). The hybrid approach yields a better overall description of the centrality-dependent balance between the evolution of radial and anisotropic flow, which particularly affects the $p_T$-spectra and differential elliptic flow of heavy particles such as protons \cite{Heinz:2011kt}.  

\subsection{Pre-equilibrium dynamics}
\label{sec2.5}

The hydrodynamic stage of ultra-relativistic heavy-ion collisions is preceded by a short ($0.2{-}1.5$\,fm/$c$) but very dense pre-equilibrium stage. During this stage the bulk of the energy density is contributed by gluon fields from the low-momentum components of the wave functions of the colliding nuclei. These fields are so intense that, in spite of weak QCD coupling $\alpha_s=g^2/(4\pi)\sim0.3<1$, interaction mean free paths are too short for a quasiparticle-based kinetic approach to make much sense. A pre-equilibrium dynamical  approach based on the non-Abelian evolution of classical gluon fields (the ``Glasma" \cite{Kovner:1995ja,Kovchegov:1997ke,Krasnitz:1998ns,Lappi:2003bi}) is probably a better starting point. 

Due to the finite size and anisotropic shape of the initial energy density distribution in the nuclear reaction zone, which fluctuates from collision to collision, any sort of interaction among the primordial QCD degrees of freedom in the fireball will cause non-vanishing radial and anisotropic flows even before the system has thermalized and viscous hydrodynamics becomes applicable. In fact, even if the system free-streams, correlations are generated between the average momenta and positions of the constituents that, when matched by the Landau prescription to a hydrodynamic form of the energy-momentum tensor, translate into non-vanishing anisotropic flow velocities on the matching surface. The hydrodynamic stage thus starts with non-zero radial flow and non-vanishing flow anisotropies,  but at the same time the pre-equilibrium evolution tends to dampen the initial spatial anisotropies and hence the anisotropic pressure gradients that drive anisotropic flow during the subsequent hydrodynamic stage. One therefore expects a certain degree of complementarity between pre- and post-equilibrium flow, resulting in reduced sensitivity of the final spectra to the starting time of the hydrodynamic stage \cite{Gale:2012rq,Vredevoogd:2008id}.      

\begin{figure}
\hspace*{-0.6cm}
\begin{minipage}{18cm}
\includegraphics[width=6cm,clip=]{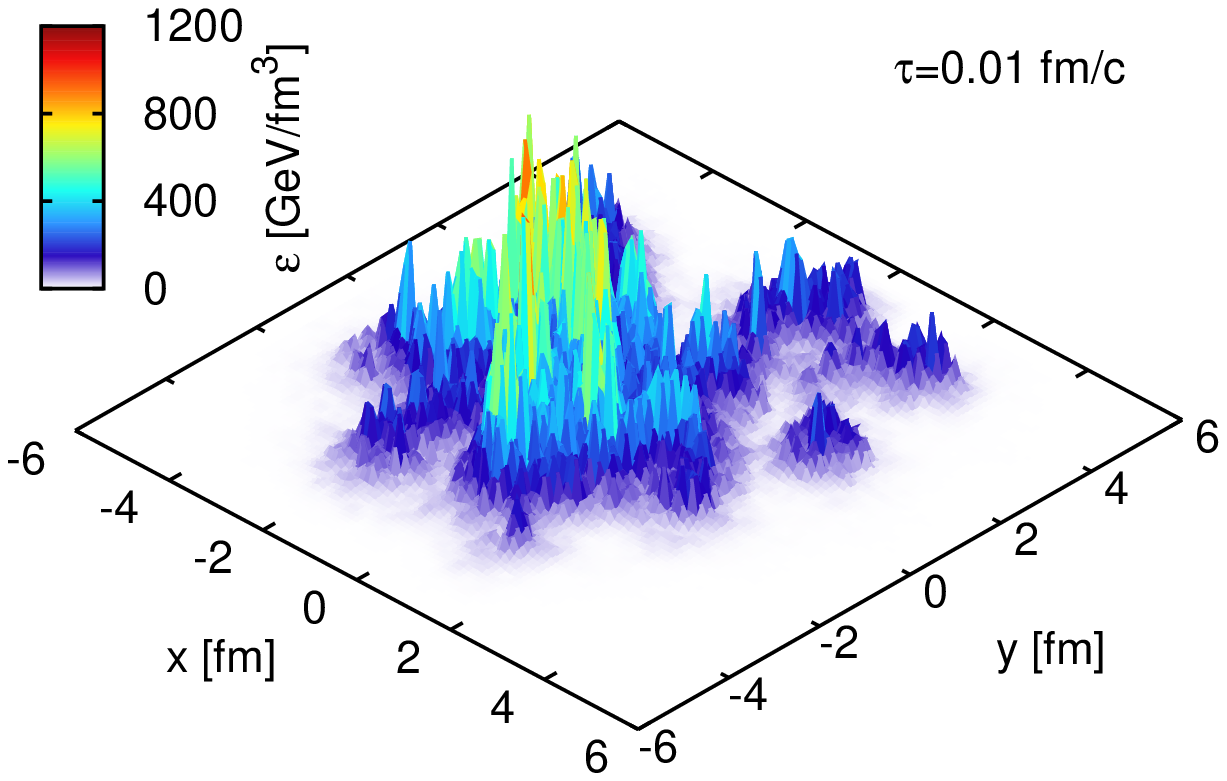}
\includegraphics[width=6cm,clip=]{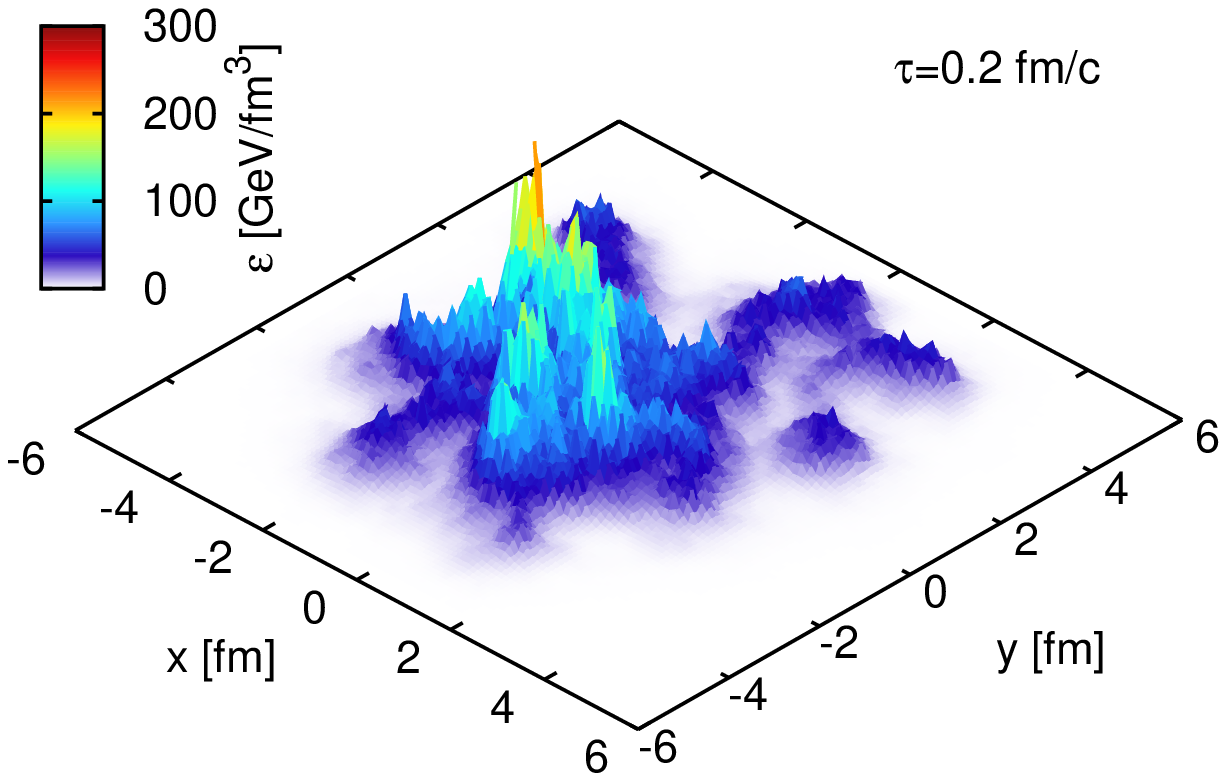}
\includegraphics[width=6cm,clip=]{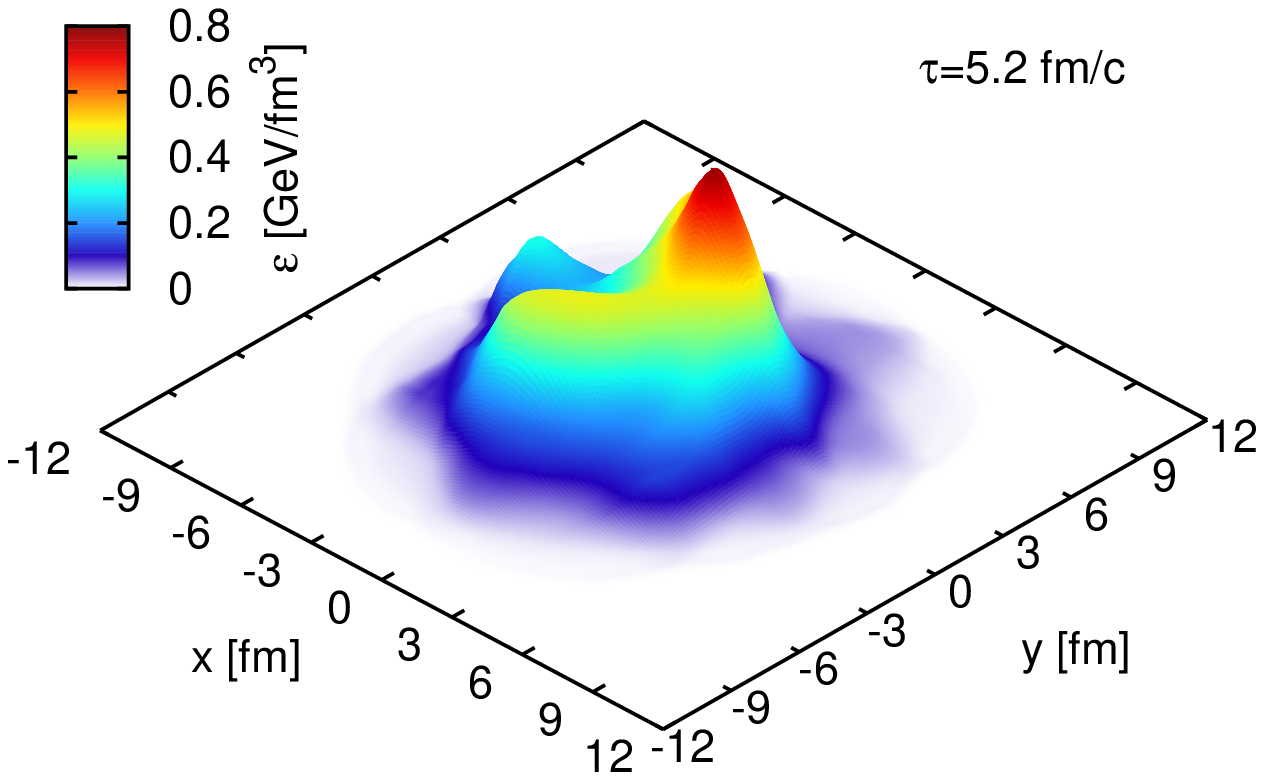}
\caption{
\label{F1}
Typical transverse energy density profiles $e(x,y)$ from the IP-Glasma model \cite{Schenke:2012wb} for a semiperipheral ($b{\,=\,}8$\,fm) Au+Au collision at $\sqrt{s}{\,=\,}200\,A$\,GeV, at times $\tau=0.01,\ 0.2,$ and 5.2\,fm/$c$. From $\tau{\,=\,}0.01$\,fm/$c$ to 0.2\,fm/$c$ the fireball evolves out of equilibrium according to the Glasma model \cite{Kovner:1995ja,Kovchegov:1997ke,Krasnitz:1998ns,Lappi:2003bi}; at $\tau{\,=\,}0.2$\,fm/$c$ the energy momentum tensor from the IP-Glasma evolution is Landau-matched to ideal fluid form (for technical reasons \cite{Gale:2012rq} the viscous pressure components are set to zero at the matching time) and henceforth evolved with viscous Israel-Stewart fluid dynamics, assuming $\eta/s{\,=\,}0.12$ for the specific shear viscosity. The pre-equilibrium Glasma evolution is seen to somewhat wash out the large initial energy density fluctuations. The subsequent viscous hydrodynamic evolution further smoothes these fluctuations. The asymmetric pressure gradients due to the prominent dipole asymmetry in the initial state of this particular event (visible as a left-right asymmetry of the density profile in the left panel) is seen to generate a dipole (``directed flow'') component in the hydrodynamic flow pattern that pushes matter towards the right during the later evolution stages.
}
\end{minipage}
\end{figure}

Model studies based on ideal fluid dynamics suggest that non-zero initial radial flow at the beginning of the hydrodynamic evolution is phenomenologically preferred by some final state observables \cite{Broniowski:2008vp}. However, faced with the conceptual difficulties of dealing with the early pre-equilibrium dynamics properly, most studies simply assumed an early hydrodynamic starting time with zero initial transverse flow velocities. With the recent discovery of the importance of initial-state density fluctuations (``hot'' and ``cold'' spots) in the early fireball this head-in-the-sand attitude became increasingly untenable.  During the last year, the dynamical IP-Glasma model \cite{Schenke:2012wb} was developed which builds on the IP-Sat (Impact Parameter dependent Saturation) model \cite{Bartels:2002cj} to generate finite deformed fluctuating initial gluon field configurations in the transverse plane, and then evolves them with classical Yang-Mills dynamics \cite{Kovner:1995ja,Kovchegov:1997ke,Krasnitz:1998ns,Lappi:2003bi}. While the lack of thermalization and of longitudinal fluctuations are still weaknesses of this model, it is the first semi-realistic approach to describing the pre-equilibrium stage {\em dynamically}, matching it consistently to the hydrodynamic stage.\footnote{It has been suggested that, since classical Yang-Mills dynamics does not lead to local thermalization, a different matching scheme \cite{Martinez:2010sc} should be used that, unlike Landau matching, does not rely on small deviations from local equilibrium. This interesting suggestion still needs to be fully worked out for fluctuating initial conditions.} {\bf Figure~\ref{F1}} shows three snapshots of the transverse energy density profile from this model.

\section{INITIAL-STATE DENSITY AND SHAPE FLUCTUATIONS}
\label{initial}

\subsection{Harmonic eccentricity and flow coefficients}
\label{sec3.1}

The development of anisotropic flow is controlled by the anisotropies in the pressure gradients which in turn depend on the shape and structure of the initial density profile. The latter can be characterized by a set of harmonic eccentricity coefficients $\varepsilon_n$ and associated angles $\Phi_n$:
\begin{equation}
\label{eq3.1}
\varepsilon_1 e^{i\Phi_1} \equiv - \frac{\int  r \, dr \, d\phi \, r^3 e^{i\phi} \, e(r,\phi)}
                                                            {\int r \, dr \, d\phi \, r^3 e(r,\phi)}, \  
\varepsilon_n e^{in\Phi_n} \equiv - \frac{\int  r \, dr \, d\phi \, r^n e^{in\phi} \, e(r,\phi)}
                                                              {\int r \, dr \, d\phi \, r^n e(r,\phi)}\  (n>1),
\end{equation}
where $e(r,\phi)$ is the initial energy density distribution in the plane transverse to the beam direction. When, for collisions between nuclei of the same species, $e$ is averaged over many events and the angle $\phi$ is measured relative to the impact parameter vector, there is a symmetry between $\phi$ and $-\phi$ as well as between $\phi$ and $\phi+\pi$, and all odd $\varepsilon_n$ coefficients vanish.

An important insight \cite{Socolowski:2004hw,Sorensen:2010zq,Alver:2010gr} has been that, due to event-by-event fluctuations of the transverse positions of the nucleons inside the colliding nuclei \cite{Miller:2003kd}, and of the gluon density profiles inside those nucleons \cite{Kovner:1995ja,Kovchegov:1997ke,Krasnitz:1998ns,Lappi:2003bi,Schenke:2012wb,Muller:2011bb,Dumitru:2012yr} (see {\bf Figure~\ref{F1}}), these symmetries do not hold in an individual collision event. Therefore, in every collision {\em all} eccentricity coefficients are usually non-zero, driving anisotropic flow components of {\em any} harmonic order whose magnitudes and directions fluctuate from event to event. The statistical distributions of $\varepsilon_n$ and $\Phi_n$ which, in a hydrodynamic picture, control the statistical distributions of the final anisotropic flows $v_n$ and their directions $\Psi_n$, are of quantum mechanical origin and depend on the internal structure of the colliding nuclei (see Sec.~\ref{sec3.3}).

The anisotropic flow coefficients $v_n$ and their associated flow angles $\Psi_n$ are defined in analogy to Eq.~(\ref{eq3.1}) as
\begin{equation}
\label{eq3.2}
  v_n e^{in\Psi_n} \equiv \langle e^{in\varphi}\rangle,\qquad
  v_n(p_T) e^{in\Psi_n(p_T)} \equiv \langle e^{in\varphi}\rangle_{p_T}, 
\end{equation}
where $\varphi$ is the azimuthal angle of the transverse momentum vector $\bm{p}_T$ of the emitted particle in the laboratory frame. Note that the $v_n$ characterize a single collision. 
In the left equation, which defines the ``$p_T$-integrated'' or ``total'' anisotropic flow, the average is over all particles of a given kind (identified by species or not) in the event. The right equation defines the ``$p_T$-differential'' flow and averages only over particles with a given transverse momentum. $v_1$, $v_2$, and $v_3$ are known as directed, elliptic, and triangular flow, respectively.

Due to the limited number of particles emitted from a single event, statistically precise measurements of $v_n$ and $\Psi_n$ can only be obtained from particle distributions that have been averaged over many events. One defines the mean flow coefficient $\bar{v}_n$ and the mean flow angle $\bar{\Psi}_n$ through a Fourier decomposition of the experimentally determined, event averaged particle distribution:
\begin{equation}
\label{eq3.3}
  \frac{d\bar{N}}{d\varphi} = \frac{\bar{N}}{2\pi} \left(1 + 2 \sum_{n=1}^{\infty} \bar{v}_n \cos(n(\varphi{-}\bar{\Psi}_n)) \right),
\end{equation}
where $\bar{N}{\,\equiv\,}\langle N\rangle$ is the mean number of particles of interest per event (charged hadrons or identified particles of a specific species). The generalization to $p_T$-differential $\bar{v}_n(p_T)$ and $\bar{\Psi}_n(p_T)$ in terms of $d{\bar{N}}/(p_T dp_T d\varphi)$ is obvious. The Fourier coefficients are given by:
\begin{equation}
\label{eq3.4}
 \bar{v}_n = \dla \cos[n(\varphi{-}\bar{\Psi}_n)]\dra,
  \;\;\; {\rm or\ equivalently} \;\;\; \bar{v}_n = \dla e^{in\varphi}\dra e^{-in\bar{\Psi}_n},
\end{equation}
where $\dla\dots\dra$ denotes the double average over all particles in an event and over all events. Note that, due to event-by-event fluctuations of the flow angle $\Psi_n$, the right expression is {\em not} identical with the average of Eq.~(\ref{eq3.2}) over collision events as is usually assumed. Observable consequences of these event-by-event fluctuations of the flow planes for generally employed experimental $v_n$ measures (to be discussed next) are only now being investigated \cite{Gardim:2012im,Qiu:2013??}.

Since the flow planes are not experimentally known, the anisotropic flow coefficients are calculated using azimuthal angular correlations between the observed particles. In the case of two particle correlations $\dla e^{in(\varphi_1{-}\varphi_2)} \dra$ the measurement is proportional to $\langle v_n^2\rangle$: under the assumption that the only azimuthal correlation between particles is due to the common correlation with the flow plane this correlator can be factorized into $\dla e^{in(\varphi_1{-}\Psi_n)}\rangle\langle e^{-in(\varphi_2{-}\Psi_n)}\dra \equiv \langle v_n^2 \rangle$. From this method, the experimentally reported anisotropic flow coefficients are therefore obtained as the root mean square value $\sqrt{\langle v_n^2 \rangle}$. Due to event-by-event fluctuations in the anisotropic flow, the event averaged $\langle v_n^k \rangle \neq \langle v_n \rangle^k$ for $k \ge 2$. We denote by $v_n\{2\} \equiv \sqrt{\langle v_n^2 \rangle}$ the anisotropic flow extracted from two-particle correlations.

In practice, not all azimuthal correlations in the data are of collective origin. Additional ``non-flow'' correlations arise from resonance decays, jet fragmentation, and Bose-Einstein correlations \cite{Voloshin:2008dg}. They can be suppressed by appropriate kinematic cuts \cite{Voloshin:2008dg} or by using multi-particle correlations known as higher-order cumulants \cite{Borghini:2001vi}. $v_n\{4\}$ and $v_n\{6\}$ denote the anisotropic flow coefficients obtained from the fourth and sixth order cumulants, respectively.

In addition to being less sensitive to non-flow contributions, the higher order cumulants, which involve higher moments of the event-by-event $v_n$ distribution, also depend differently on the variance $\sigma_{v_n}$ of that distribution. If $\sigma_{v_n}{\,\ll\,}\langle v_n \rangle$, one finds up to order $\bigl(\sigma_{v_n}/\langle v_n\rangle\bigr)^2$ 
\begin{equation}
\label{eq3.5}
    v_n\{2\} \approx \langle v_n \rangle + \frac{1}{2} \frac{\sigma_{v_n}^2}{\langle v_n \rangle} \;\;\; {\rm and} \;\;\;
    v_n\{4\} \approx \langle v_n \rangle - \frac{1}{2} \frac{\sigma_{v_n}^2}{\langle v_n \rangle}.
    \label{fluctuations}
\end{equation}
This illustrates that the difference between $v_{n}\{2\}$ and $v_{n}\{4\}$ is sensitive to the width of the $v_n$ distribution, in addition to non-flow effects. After correcting $v_n\{2\}$ for non-flow \cite{Ollitrault:2009ie}, this difference can be used to estimate the variance of the event-by-event flow fluctuations (see Sec.~\ref{sec6.4}).

\subsection{Centrality classes}
\label{sec3.2}

Heavy ions are extended objects, and the system created in a head-on collision is different from that in a peripheral collision. We therefore categorize nuclear collisions by their centrality, parametrized by the impact parameter $\bm{b}$ which is, however, not a direct observable. Experimentally, the collision centrality can be inferred from the number of produced hadrons, if one assumes that this multiplicity is a monotonic function of $\bm{b}$. Knowing which fraction of the total hadronic cross section is observed in the experiment one can divide the measured distribution of produced particles in centrally classes corresponding to the percentile of total hadronic cross section.

In addition to impact parameter or fraction of total hadronic cross section, one can also use the so-called number of wounded nucleons or the number of binary nucleon-nucleon collisions to characterize the collision centrality. These quantities are defined within the Glauber model \cite{Miller:2007ri} and are stochastic functions of the impact parameter. Phenomenologically it is found that soft particle production scales roughly with the number of participating nucleons whereas hard processes scale with the number of binary collisions. 

Anisotropic flow is not measured in a single event but in a centrality class. 
Therefore, event-by-event fluctuations due to impact parameter fluctuations within a centrality class will add to the initial-state fluctuations mentioned in Sec.~\ref{sec3.1} (and further discussed in the following subsection) to determine the spectrum of final-state flow ($v_n$ and $\Psi_n$) fluctuations.

\subsection{Models for the primordial fluctuation power spectrum}
\label{sec3.3}

Various theoretical approaches have been used to model the initial energy and entropy density profiles. The most common approach is the Monte Carlo (MC) Glauber model~\cite{Miller:2007ri}. In this model the positions of the nucleons inside the two colliding nuclei are sampled according to the measured nuclear density distribution, accounting for their finite size \cite{Hirano:2009ah}. The sampling procedure introduces event-by-event fluctuations of the nucleon positions, representing the quantum mechanical fluctuations of the outcome of a position measurement on the nucleons whose probability distribution is given by the (smooth) nuclear ground state wave function. The nucleons travel on straight-line trajectories and collide if their distance in the transverse plane is smaller than the radius corresponding to the total inelastic nucleon-nucleon cross-section. It is assumed that this radius is independent of the number of interactions the nucleons already had. Nucleons in the target and projectile that have had at least one interaction are called participants or wounded nucleons. Soft particle production is assumed to be proportional to the number density of wounded nucleons whereas hard (high-$p_T$) processes are taken to scale with the number density of binary nucleon-nucleon collisions. The initial entropy or energy density profile is typically taken proportional to a linear combination of the wounded nucleon and binary collision densities which are strongly fluctuating in the transverse plane and from event to event.

In the Monte Carlo Kharzeev-Levin-Nardi (MC-KLN) model~\cite{Drescher:2006ca} the entropy production is determined by the initial gluon production, calculated by perturbative merging of two gluons from the projectile and target nuclei where (in the spirit of Color Glass Condensate ideas \cite{KL2012}) the ($p_T$-unintegrated) gluon structure functions are parametrized by a position dependent gluon saturation momentum $Q_s$ \cite{Kharzeev:2001yq}. $Q_s(\bm{r}_\perp)$ is computed from the longitudinally projected density of wounded nucleons whose positions are sampled as in the MC-Glauber model.

The MC-Glauber and MC-KLN models do not account for fluctuations of the gluon fields inside the colliding nucleons which are characterized by a correlation length $\sim 1/Q_s$ in the transverse plane \cite{Muller:2011bb}. For this reason, they cannot reproduce the measured multiplicity fluctuations in $pp$ collisions \cite{Dumitru:2012yr}. Gluon field fluctuations can be imprinted on the MC-Glauber or MC-KLN profiles {\it a posteriori} using an algorithm developed in \cite{Moreland:2012qw}. In the IP-Glasma model, described at the end of Sec.~\ref{sec2.5}, gluon field fluctuations are imprinted on the IP-Sat model \cite{Bartels:2002cj} using ideas from the Color Glass Condensate/Glasma theory \cite{Kovner:1995ja,Kovchegov:1997ke,KL2012} and then evolved using classical Yang-Mills dynamics \cite{Krasnitz:1998ns,Lappi:2003bi} to a matching surface after which hydrodynamics takes over.

{\tt DIPSY} is a Monte Carlo event generator based on gluon radiation from colored dipoles (via dipole splitting) using BFKL evolution \cite{Flensburg:2011kk}. The nucleons in the colliding nuclei are described by a triangle of color dipoles whose position is sampled from a Woods-Saxon distribution just as in the MC-Glauber and MC-KLN models. Gluon density fluctuations arise from the Monte Carlo sampling of the radiative gluon shower.

The Monte Carlo approach used in these models allows for an event-by-event calculation of the eccentricities $\varepsilon_n$ and thus for the determination of the higher order moments of the $\varepsilon_n$ distributions. Because of the non-zero widths of these distributions, the measured final flow power spectrum obtained from a given moment of the $v_n$ distribution, say $v_n\{k\}$, should be compared with the initial fluctuation power spectrum $\varepsilon_n\{k\}$ obtained from the analogous moment of the $\varepsilon_n$ distribution \cite{Voloshin:2008dg}.

%
\begin{figure}[thb]
 \begin{center}
   \includegraphics[width=0.7\textwidth]{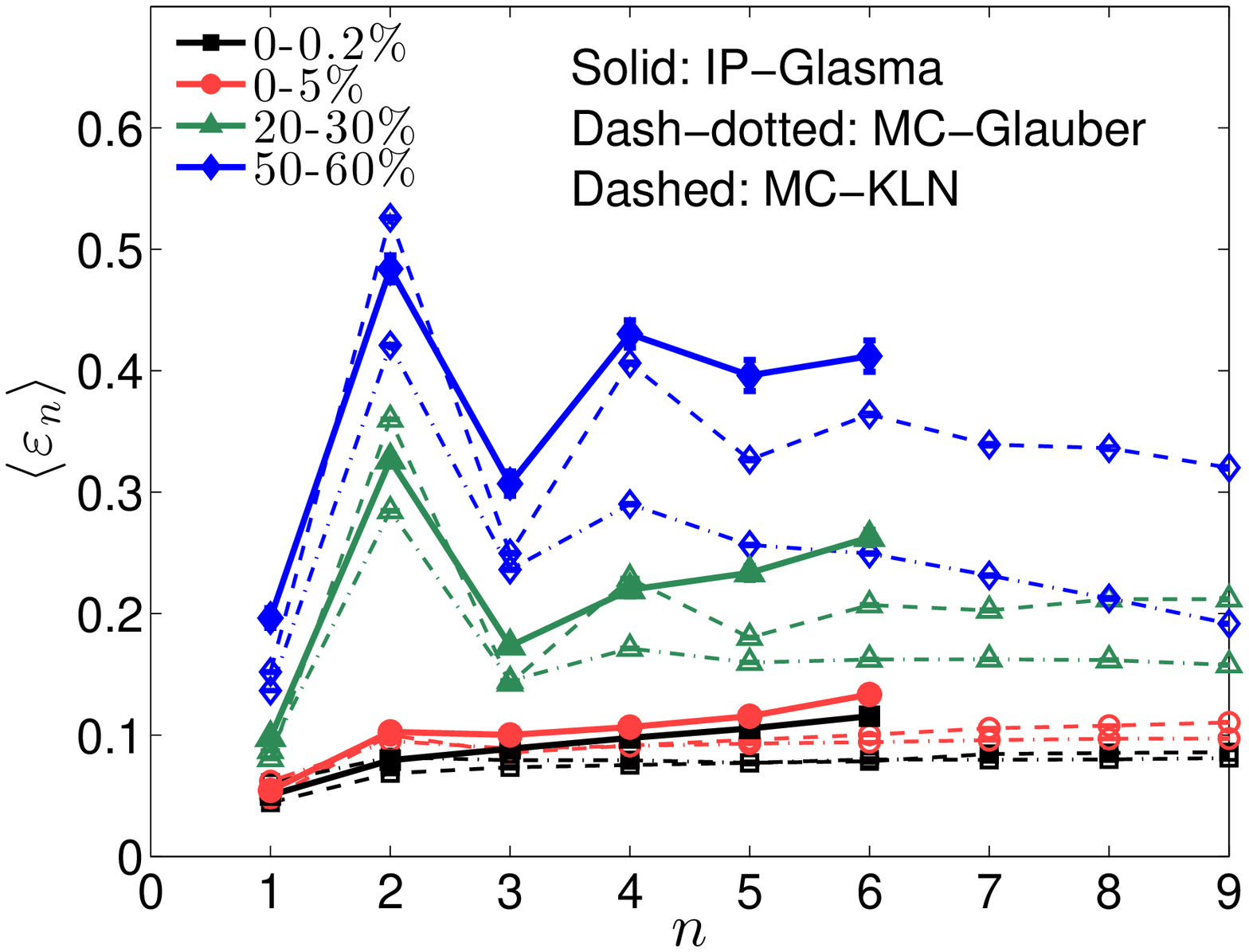}
 \end{center}
 \caption{
\label{F2}
Primordial fluctuation power spectrum of the Little Bangs created in 2.76\,$A$\,TeV Pb+Pb collisions of different centralities, from three different initial-state models (IP-Glasma, MC-Glauber, MC-KLN).  
 }
\end{figure}
%
{\bf Figure~\ref{F2}} shows the $\langle\varepsilon_n\rangle$ power spectrum of the initial energy density for the MC-Glauber, MC-KLN and IP-Glasma models, for four centrality classes. In the most central collisions, $0-0.2$\%, the $\varepsilon_n$ are entirely due to fluctuations, and therefore all $\langle\varepsilon_n\rangle$ have roughly equal magnitudes. Proceeding to less central collisions, the nuclear overlap region develops a pronounced elliptic geometric deformation which increases $\langle\varepsilon_2\rangle$ (and, to some extent, also $\langle\varepsilon_4\rangle$ and $\langle\varepsilon_6\rangle$) much more strongly than the odd eccentricity coefficients which remain fluctuation dominated. Importantly, {\bf Figure~\ref{F2}} reveals a significant model dependence of the $\langle\varepsilon_n\rangle$ spectrum. This implies on the one hand that the extraction of QGP transport coefficients (see Sec.~\ref{sec4.2}) is complicated by uncertainties related to initial model ambiguities, but on the other hand that, by constraining the theory with a sufficiently large set of anisotropic flow and correlation measurements (see Chap.~\ref{vn}), one can discriminate phenomenologically between different initial state models and identify the one that produces the correct initial fluctuation power spectrum.

The different initial state models also produce different correlations between the ``participant plane angles'' $\Phi_n$ associated with $\varepsilon_n$. For the MC-Glauber and MC-KLN models these were studied in \cite{Bhalerao:2011yg,Jia:2012ma,Qiu:2012uy}. The evolution of these initial-state participant plane correlations through non-linear viscous hydrodynamics into final-state flow-plane correlations \cite{Qiu:2012uy,Teaney:2012ke} and the experimental measurement of the latter \cite{ALICE:2011ab,Jia:2012sa} will be discussed in Sec.~\ref{sec6.3}.

\section{HYDRODYNAMIC RESPONSE TO INITIAL-STATE FLUCTUATIONS}
\label{response}

\subsection{Viscous effects on radial and anisotropic flow}
\label{sec4.1}

The viscosity of a fluid is related to its ability to return to local thermal equilibrium after being driven away from equilibrium by gradients in its macroscopic flow pattern. When the corresponding relaxation times approach zero, viscous effects disappear. Small viscosities are therefore related to short relaxation times, i.e. strong interactions among the microscopic constituents.

For non-zero relaxation times, viscous effects cause the microscopic momentum distribution in the local fluid rest frame to deviate from its exponential local equilibrium form. {\em Bulk viscosity} causes locally isotropic deviations from equilibrium, adding a diagonal contribution, the bulk viscous pressure $\Pi\,\delta^{ij}$, to the stress tensor $T^{ij}$  in the local rest frame. In Navier-Stokes approximation (which ignores retardation) it is proportional to the scalar fluid expansion rate $\theta$ at the location of the fluid cell, $\Pi(x)=-\zeta\theta(x)$. For an expanding fluid this is negative, i.e. the bulk viscous pressure counteracts the expansion. In an isotropically expanding fireball, bulk viscosity reduces the radial acceleration and thus inhibits the buildup of radial flow.

{\em Shear viscosity} causes locally anisotropic deviations from equilibrium, resulting in an anisotropic contribution to the local rest frame stress tensor. This shear viscous pressure is driven by shear flow and acts against flow anisotropies. In relativistic heavy-ion collisions, due to approximate boost-invariance along the beam direction of the physical processes that generate the quark-gluon plasma, the initial expansion rate is highly anisotropic and much larger along the beam direction than transverse to it. Transverse flow builds only later, in response to transverse pressure gradients in the initial state. The main effect of shear viscosity is that it tries to equalize the expansion rates along different directions, by building up a shear viscous pressure tensor $\pi^{ij}$ in the local rest frame that reduces the longitudinal and increases the transverse pressure. Less work is done by longitudinal pressure, causing the QGP to cool less rapidly (at least initially when cooling is dominated by longitudinal expansion) but simultaneously increasing the build-up of transverse flow. Transverse anisotropies of the initial fireball geometry, reflected in anisotropic transverse pressure gradients, generate anisotropies in the developing transverse flow; shear viscosity reduces these flow anisotropies, i.e. it degrades the medium's ability to convert initial transverse pressure anisotropies into final transverse flow anisotropies.  

Anisotropic transverse flow influences the shape of the transverse momentum spectra of the finally emitted particles through a direction-dependent blue-shift factor. By reducing the azimuthally symmetric component of transverse flow, a.k.a. radial flow, bulk viscosity leads to steeper $p_T$-spectra; by boosting radial flow, shear viscosity renders them flatter. Numerically, it was found \cite{Song:2009rh} that in heavy-ion collisions shear viscous effects dominate over bulk viscous ones by about a factor 5--10. Experimentally, they are not easy to separate since both types of effects modify the slopes of the $p_T$-spectra as well as the anisotropic flow coefficients discussed below. A clear strategy for systematically isolating shear from bulk viscous effects still needs to be formulated. Lacking such a strategy, most researchers presently allow themselves to be guided by available theoretical studies and interpret anisotropic flow measurements entirely in terms of shear viscosity, i.e. they ignore bulk viscosity in their theoretical models. 
 
\subsection{Extracting the QGP shear viscosity from experimental data}
\label{sec4.2}

Shear viscosity affects the $p_T$-spectra of the finally emitted particles in two distinct ways: (i) it increases the magnitude and decreases the anisotropies of the hydrodynamically generated transverse flow, and (ii) it causes a deviation $\delta f$ of the final phase-space distribution $f(x,p)$ from its isotropic local equilibrium form $f_0(x,p)$: $f(x,p)=f_0(x,p)+\delta f(x,p)$. The first of these two effects reflects the amount of shear viscosity over the entire expansion history of the fireball; the second effect, $\delta f$, is only sensitive to the shear viscosity at the final conversion to hadrons (assuming, as required for the applicability of viscous hydrodynamics, that microscopic relaxation rates are much larger than the macroscopic expansion rate). $\delta f$ increases with transverse momentum roughly as $p_T^\alpha$ where $\alpha$ ranges between 1 and 2 \cite{Teaney:2003kp}. $\delta f$ is thus small at low $p_T \lapp 5\,T$ (where $T\sim 100{-}160$\,MeV is the decoupling temperature), but at larger $p_T\gapp 2-3$\,GeV it becomes so big \cite{Song:2007fn} that the near-equilibrium expansion (and thus the hydrodynamic prediction for the $p_T$-distribution) can no longer be trusted. Being predominantly interested in the shear viscosity $(\eta/s)_\mathrm{QGP}$ of the early QGP phase, one would like to minimize in the analysis the effects from $\delta f$ which reflect only the late hadronic shear viscosity. This can be achieved by studying $p_T$-integrated observables which are dominated by transverse momenta $p_T \lapp 5\,T$ where $\delta f$ is small.  
   
A second consideration is that the hydrodynamically generated transverse momentum anisotropy is distributed in the measured final state over a large number of different hadronic species. Since heavier particles are affected by radial flow more strongly than lighter particles \cite{SSHPRC93}, resulting in flatter $p_T$-spectra at low-$p_T$ where most final hadrons find themselves, the chemical composition of the final state, in particular the light-to-heavy particle ratios, control where in $p_T$ the hydrodynamic momentum anisotropy ends up: For heavy hadrons, radial flow pushes the flow anisotropies to larger $p_T$; at low $p_T$ the anisotropic flow is dominated by light hadron species. Extracting the shear viscosity $\eta/s$ from the $p_T$-dependent anisotropic flow coefficients thus introduces an undesirable fragility related to the precision with which the theoretical model describes the chemical composition of the fireball at freeze-out and the shapes of the different hadronic $p_T$-spectra that control the distribution in $p_T$ of the hydrodynamically generated momentum anisotropy. A more robust extraction of $\eta/s$ uses the azimuthal anisotropy coefficients of the $p_T$-integrated angular distributions, summed over all hadronic species \cite{Heinz:2005zg,Teaney:2009qa,Song:2010mg,Song:2011hk} (the so-called ``charged hadron $v_n$''), and later checks the $p_T$-differential flow coefficients $v_n(p_T)$, for all charged hadrons together and for individual hadron species separately, as additional tests whether the model also correctly describes the relative hadron abundances and their $p_T$-distributions in the final freeze-out stage.

\begin{figure}[h]
\includegraphics[width=\linewidth,clip=]{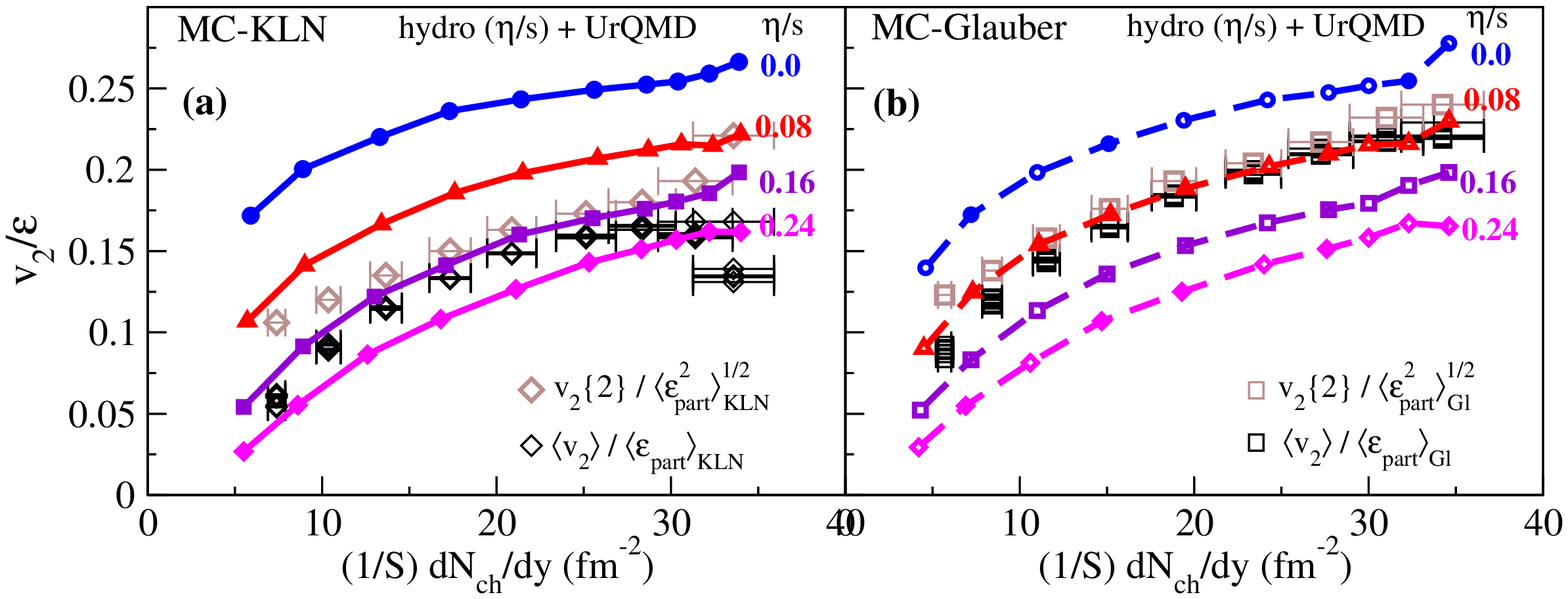}
\caption{\label{F3} 
The eccentricity-scaled integrated elliptic flow of all charged hadrons, $v_2^\mathrm{ch}(\eta/s)/\varepsilon$, as a function of total charged hadron multiplicity density per unit overlap area, $(1/S)(dN_\mathrm{ch}/dy)$. The experimental data points show two measures for the elliptic flow ($\langle v_2\rangle$ \cite{Ollitrault:2009ie} and $v_2\{2\}$ \cite{Adams:2004bi}) from 200\,$A$\,GeV Au-Au collisions at RHIC, measured by the STAR Collaboration. Both panels use the same sets of data, but use different normalization average initial eccentricities $\langle \varepsilon\rangle$ and overlap areas $\langle S\rangle$ to normalize the vertical and horizontal axes, obtained from the Monte-Carlo-Glauber \cite{Miller:2007ri} and Monte-Carlo Kharzeev-Levin-Nardi (MC-KLN \cite{Kharzeev:2001yq,Drescher:2006ca}) models. The theoretical curves were computed with the {\tt VISHNU} model \cite{Song:2010aq,Song:2010mg}, for different (temperature-independent) choices of the specific QGP shear viscosity $(\eta/s)_\mathrm{QGP}$.
}
\end{figure}

{\bf Figure~\ref{F3}} shows an attempt of such an extraction of the specific QGP shear viscosity from charged hadron elliptic flow ($v_2$) data collected at RHIC. The plot shows $v_2$ scaled by the initial fireball {\em ellipticity} $\varepsilon{\,\equiv\,}\varepsilon_2$ ($\varepsilon$ characterizes the elliptical spatial deformation of the initial pressure gradients that drive the hydrodynamic expansion of the fireball) as a function of collision centrality, represented by the charged hadron multiplicity density per unit overlap area $(1/S)(dN_\mathrm{ch}/dy)$ (the most central collisions have the largest multiplicity density). The theoretical curves were obtained with the {\tt VISHNU} hybrid model \cite{Song:2010aq} with a temperature-independent specific QGP shear viscosity $(\eta/s)_\mathrm{QGP}$. The model results show a monotonic dependence on $(\eta/s)_\mathrm{QGP}$ and nicely reproduce the shape of the measured centrality dependence. By normalizing $v_2$ by $\varepsilon_2$, the theoretical curves become insensitive to the experimental method used for measuring $v_2$ as long as the same procedure used in determining $v_2$ is also used to compute the initial ellipticity $\varepsilon_2$ \cite{Song:2010mg,Song:2011hk}. However, the magnitude of $\varepsilon_2$ differs by about 20\% between the two initial-state models studied in the left and right panels of {\bf Figure~\ref{F3}}, shifting the normalized experimental data up or down relative to the theoretical curves by similar amounts. As a result, the $(\eta/s)_\mathrm{QGP}$ value extracted from the comparison is uncertain by a factor $2{-}2.5$ and depends on the specific choice of the initialization model. In Sec.~\ref{vn} we review how this model ambiguity can be resolved by using experimental information from all flow harmonics $v_n$ and their systematic dependences on collision energy and centrality and on system size.

\section{TRANSVERSE MOMENTUM SPECTRA AND RADIAL FLOW}

\subsection{Radial flow systematics at RHIC and LHC energies}
\label{sec5.1}

\begin{figure}
\includegraphics[width=\linewidth,clip=]{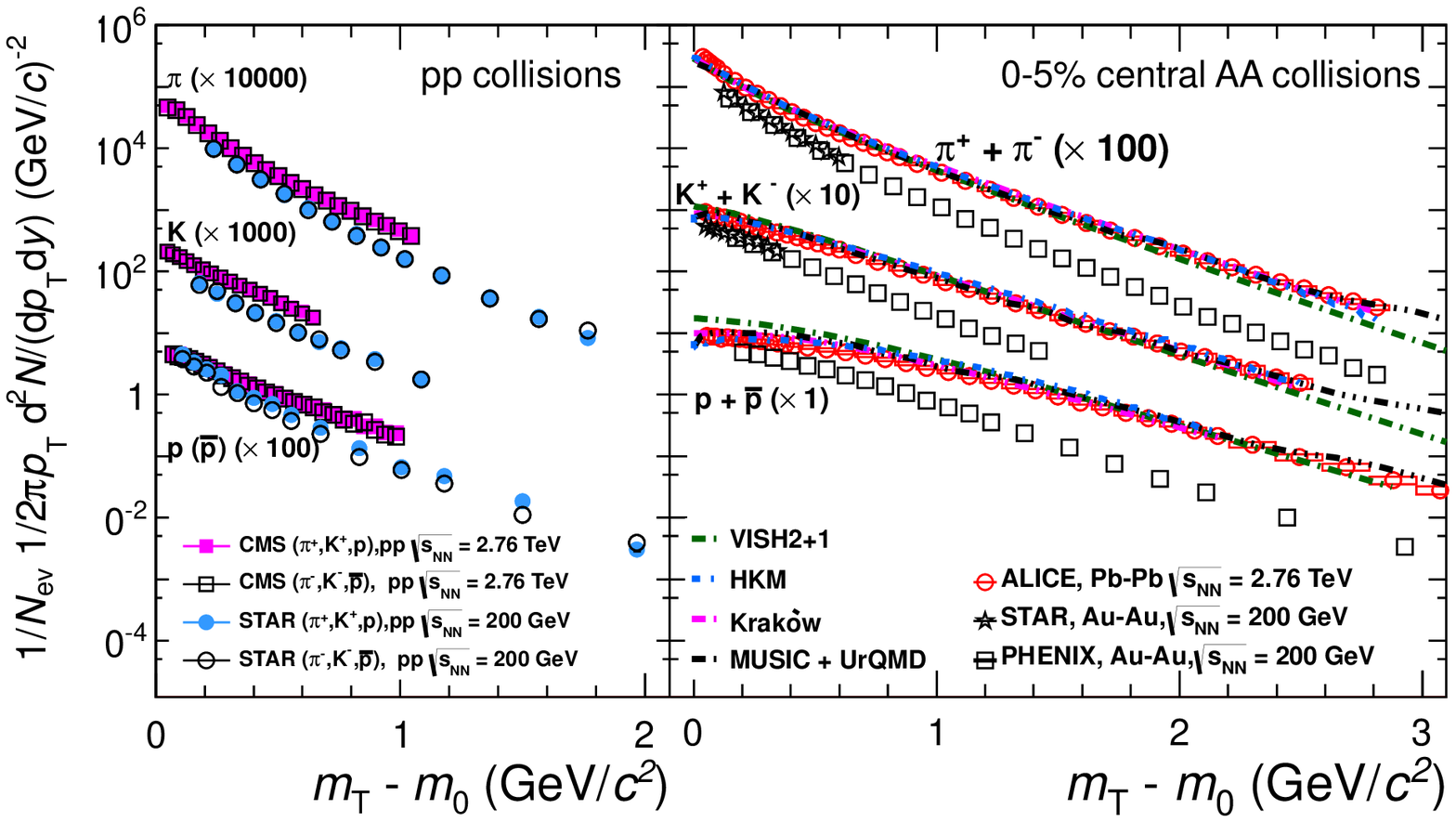}
\caption{
\label{F4} 
Transverse mass ($m_T=\sqrt{m^2+p_T^2}$) distributions of pions, kaons and protons from $pp$ and Au+Au resp. Pb+Pb collisions at RHIC and LHC, at the collision energies indicated in the plots. The data (symbols) are from Refs.~\cite{Adams:2003qm,Chatrchyan:2012qb} ($pp$ collisions at RHIC and LHC), Refs.~\cite{Adler:2003cb,Abelev:2008ab} (Au+Au collisions at RHIC), and \cite{:2012iu} (Pb+Pb collisions at the LHC). The models (lines) are described in the text.
}
\end{figure}

Radial flow causes a blueshift of the transverse momenta of the finally emitted hadrons, leading  to flatter $p_T$ and $m_T$ distributions, especially at low $p_T$ where non-relativistic kinematics allows to express $p_T\approx p_T^\mathrm{th} + m \langle v_T\rangle$ as the sum of a thermal contribution $p_T^\mathrm{th}$ (which depends only on the decoupling temperature $T_\mathrm{kin}$ and is independent of the hadron mass $m$) and a collective flow component with average flow velocity $\langle v_T\rangle$ which is proportional to the hadron mass \cite{SSHPRC93}. This is shown in {\bf Figure~\ref{F4}}: Due to collective flow, the heavy-ion collision spectra in the right panel are flatter than the $pp$ spectra in the left panel, and due to $\sim10\%$ stronger flow at the higher collision energy, they are flatter at the LHC than at RHIC. For heavy hadrons such as protons, the radial flow generates a ``shoulder" in the spectrum at low $p_T$ which is more pronounced at LHC than at RHIC energies. In contrast, the soft parts ($p_T<2.5$\,GeV/$c$) of the $pp$ spectra do not feature the flow-induced splitting between kaon and proton slope parameters that characterizes the heavy-ion collision spectra.\footnote{The different slopes of pion and kaon spectra in $pp$ collisions have a different origin, unrelated to flow: a large fraction of the pions in the final state arises from the decay of unstable heavier resonances, and these decay pions accumulate preferentially at low $p_T$, due to kinematic constraints \cite{SKH90}. The larger inverse slope of the $pp$ collision spectra at the higher LHC energy is an initial state effect, arising (among other factors) from a higher gluon density probed at LHC energies compared to RHIC \cite{Tribedy:2010ab}.}

\begin{figure}
\begin{center}
\includegraphics[width=0.7\textwidth,clip=]{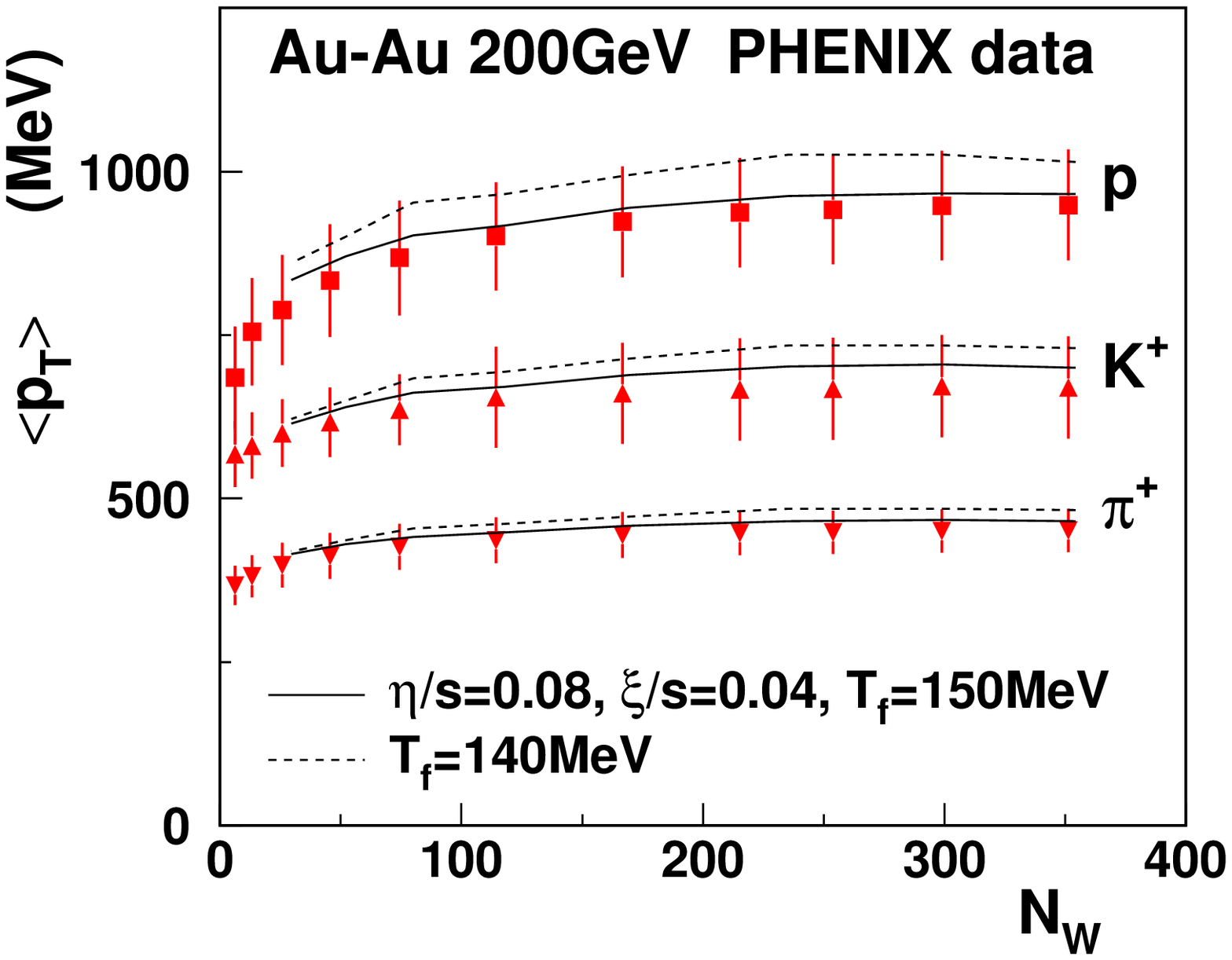}
\caption{
\label{F5} 
Average transverse momentum vs. number of wounded nucleons $N_w$, for pions, kaons and protons from 200\,$A$\,GeV Au+Au collisions. Experimental data from the PHENIX Collaboration are compared with (3+1)-dimensional event-by-event viscous hydrodynamic simulations with constant specific shear and bulk viscosities as indicated, for two different decoupling temperatures $T_\mathrm{f}\equiv T_\mathrm{kin}$. Figure taken with permission from Ref.~\cite{Bozek:2012fw}.
}
\end{center}
\end{figure}

Blast-wave model fits \cite{SSHPRC93} of the hadron spectra in terms of two parameters, the decoupling temperature $T_\mathrm{kin}$ and the average transverse flow $\langle v_T\rangle$ at kinetic decoupling, show that $\langle v_T\rangle$ increases while $T_\mathrm{kin}$ decreases with growing collision energy and as the collisions become more central (\cite{Arsene:2004fa,Kumar} and personal communication by L. Kumar). This is consistent with expectations from a kinetic theory description of the decoupling process \cite{Heinz:1999kb}. Hydrodynamic simulations show that the stronger radial flow at higher energies and in more central collisions is mostly a consequence of higher initial fireball densities, leading to a longer fireball lifetime. Stronger flow increases the mean $p_T$ of the emitted hadrons while a lower decoupling temperature decreases it. {\bf Figure~\ref{F5}} shows that, at fixed collision energy, $\langle p_T\rangle$ grows as the collisions become more central; a similar analysis for central collisions as a function of collision energy shows that $\langle p_T\rangle$ also grows with collision energy. The positive effect on $\langle p_T\rangle$ from increasing radial flow thus dominates over the negative effect from the accompanying decrease of $T_\mathrm{kin}$.

The mean $p_T$ of charged hadrons increases with increasing shear viscosity but decreases with increasing bulk viscosity. For Glauber model initial conditions it is also affected by the width $w$ of the Gaussian smearing profile \cite{Bozek:2012fw,Holopainen:2010gz}: increasing $w$ produces smoother initial density profiles and reduces the radial flow and thus $\langle p_T\rangle$. Lowering the decoupling temperature $T_\mathrm{kin}$ increases the fireball lifetime and thus the radial flow and $\langle p_T\rangle$ \cite{Bozek:2012fw}.

\begin{figure}
\includegraphics[width=\linewidth,clip=]{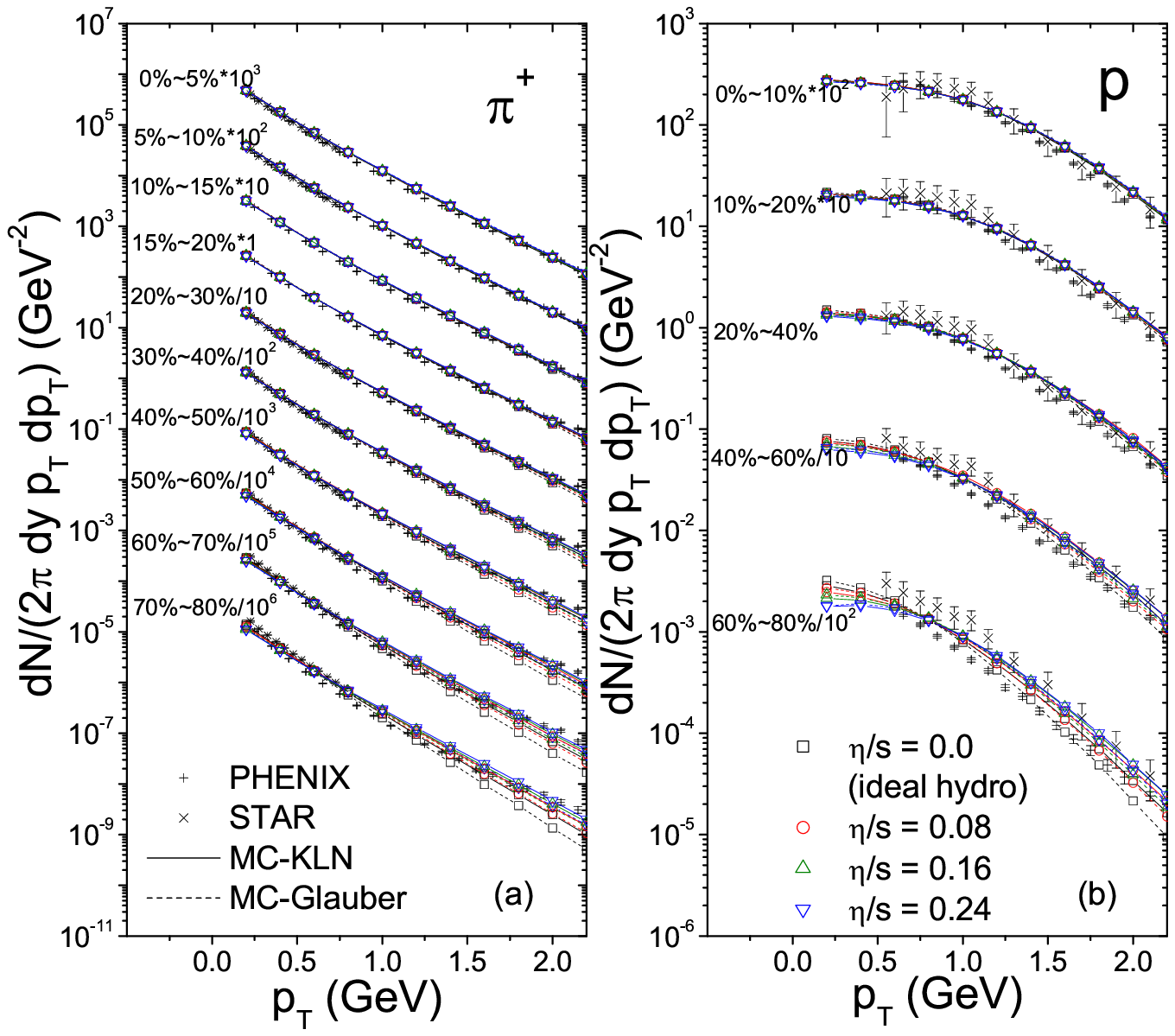}
\caption{
\label{F6} 
(a) Pion and (b) proton $p_T$ spectra from 200\,$A$\,GeV Au+Au collisions at different collision centralities, compared with viscous hydrodynamic simulations \cite{Song:2010mg}.   
}
\end{figure}

{\bf Figures~\ref{F4}} and {\bf \ref{F6}} show $p_T$ spectra for pions, kaons and protons from Au+Au and Pb+Pb collisions at RHIC and LHC. As seen in the right panel in {\bf Figure~\ref{F4}}, hydrodynamic models with ({\tt HKM} \cite{Akkelin:2008eh} and {\tt MUSIC{+}UrQMD} \cite{Ryu:2012at}) and without a hadronic cascade afterburner ({\tt VISH2{+}1} \cite{Shen:2011eg} and {\tt Krakow} \cite{Bozek:2012qs}) describe the data quite well; hydrodynamic models without hadronic afterburner that implement chemical freeze-out at $T_\mathrm{chem}{\,\approx\,}T_c$ overpredict the proton yields, by ignoring baryon-antibaryon annihilation in the hadronic phase, but still reproduce the shape of the spectra. As seen in the left panels of {\bf Figure~\ref{F6}}, the azimuthally averaged $p_T$ spectra cannot distinguish by themselves between different initial conditions and different values of the QGP shear viscosity: While larger $\eta/s$ values cause stronger radial flow, this can be compensated for by assuming longer thermalization times and a corresponding later start of the hydrodynamic evolution stage \cite{Song:2010mg} which shortens the time available for building radial flow. A determination of the QGP transport coefficients thus requires the simultaneous investigation of the azimuthally averaged $p_T$-spectra together with their azimuthal anisotropies (see Sec.~\ref{vn}).

\subsection{Radial flow fluctuations}
\label{flowflucs}

\begin{figure}
\begin{center}
\includegraphics[width=0.4\linewidth,clip=]{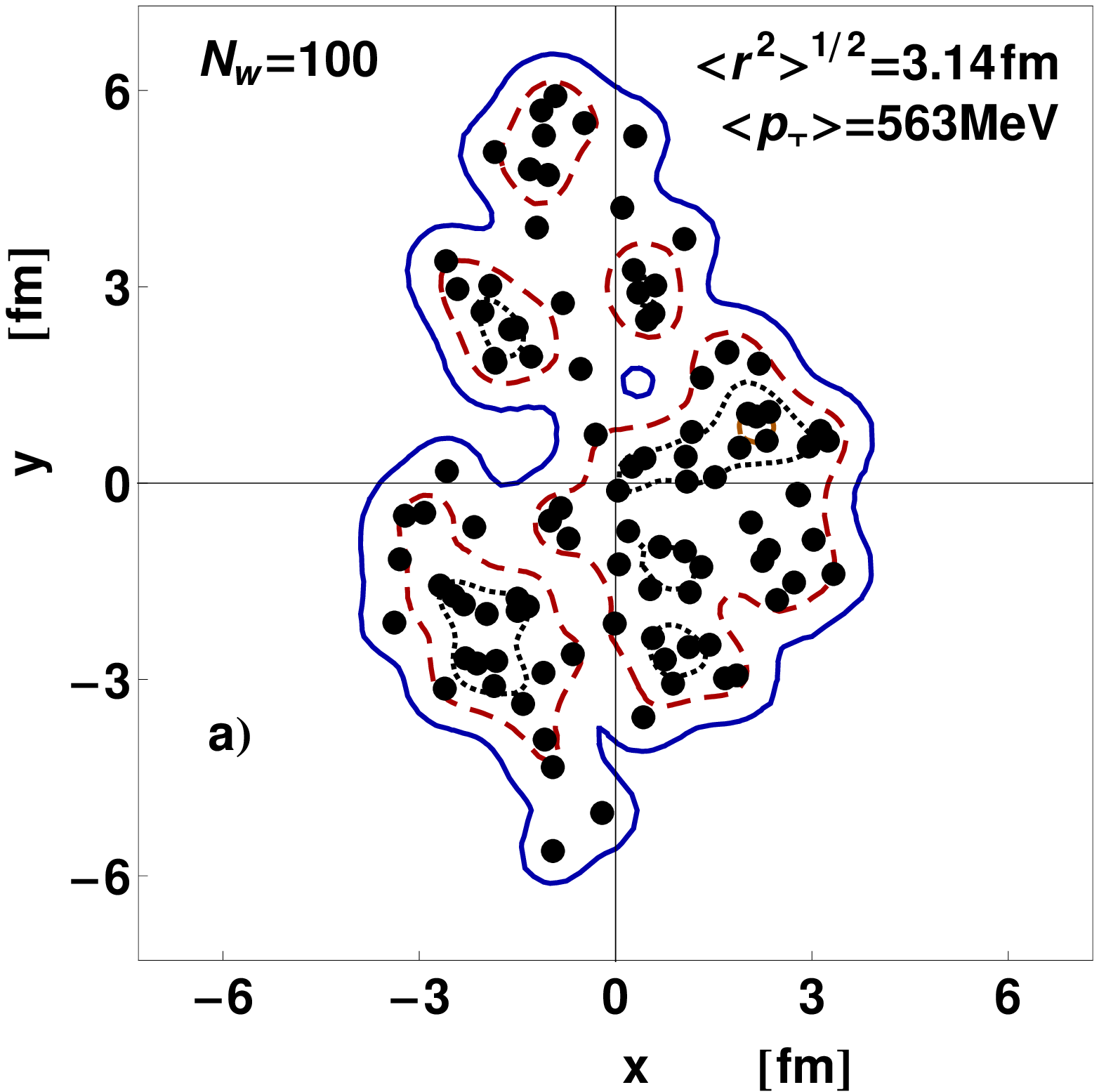}
\includegraphics[width=0.4\linewidth,clip=]{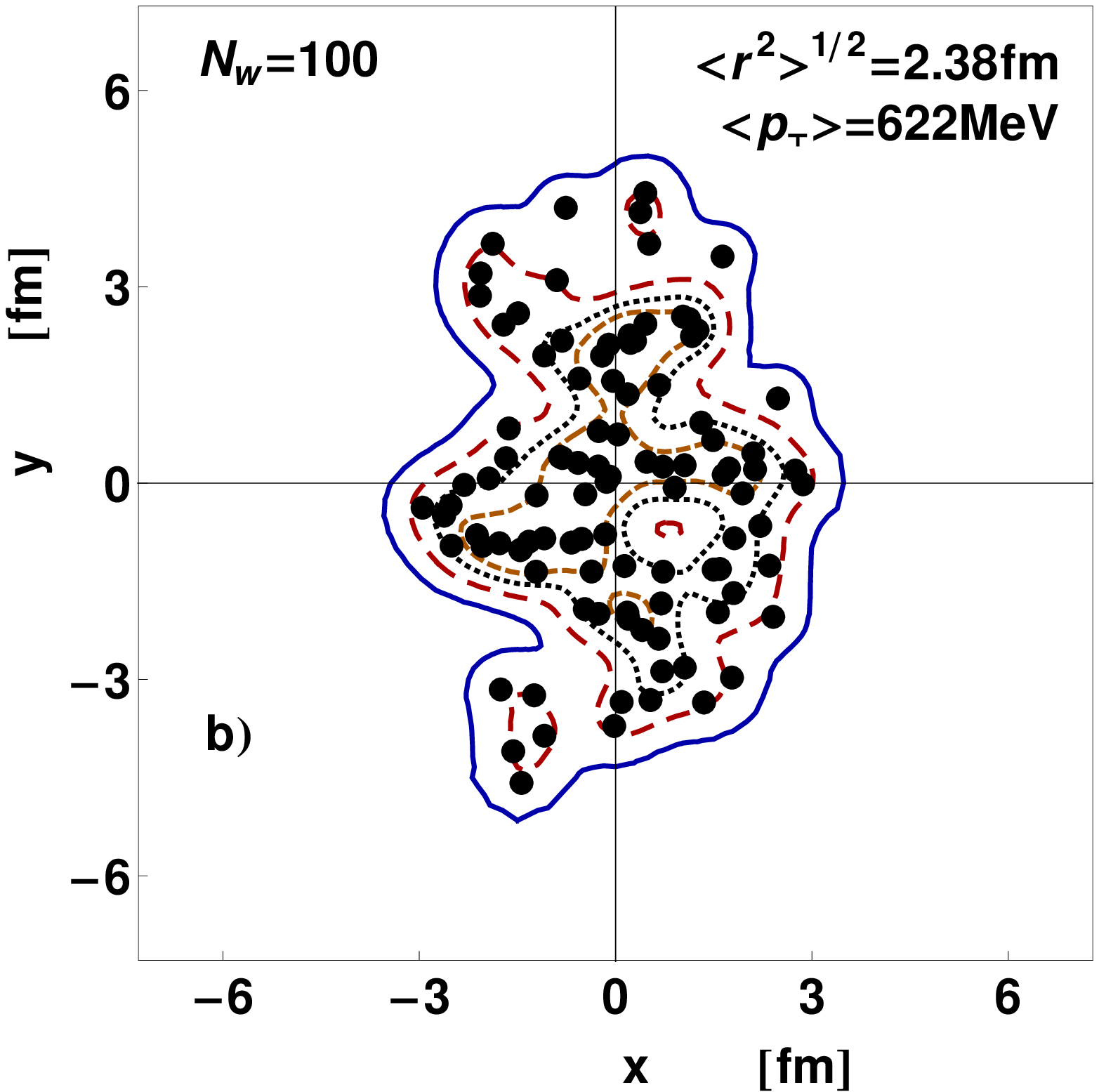}
\caption{
\label{F7} 
Two typical wounded nucleon configurations in the transverse plane 
for Au+Au collisions with $N_w=100$ wounded nucleons. Lines indicate contours of constant entropy density $s$ after smearing the wounded nucleons with Gaussian distributions. 
Figure taken with permission from Ref.~\cite{Bozek:2012fw}.
}
\end{center}
\end{figure}

Initial-state fluctuations affect not only the shape but also the size of the initial fireball \cite{Bozek:2012fw}. {\bf Figure~\ref{F7}} shows initial wounded nucleon distributions for two Au+Au collisions with equal numbers of wounded nucleons ($N_w=100$), but rms radii that differ by more than 40\% (3.14 and 2.38\,fm, respectively \cite{Bozek:2012fw}). The larger pressure gradients associated with the more compact configuration lead, through hydrodynamic evolution, to a 10\% larger mean $p_T$ of the finally observed charged hadrons (622 vs. 563\,MeV/$c$, respectively). The normalized variance $\sigma_r/\langle r\rangle$ of the fluctuations in the initial fireball radius increases from about $2{-}3\%$ in central Au+Au collisions to more than 15\% in peripheral collisions \cite{Bozek:2012fw}. Event-by-event viscous hydrodynamic evolution of these fluctuating initial states allows to compute the covariance of the resulting final state $p_T$ fluctuations. For Glauber initial conditions evolved with $\eta/s=0.08$ and a hadronic bulk viscosity of $\zeta/s=0.04$, the theoretical predictions \cite{Bozek:2012fw} compare well with available experimental data from the PHENIX and STAR Collaborations \cite{Adler:2003xq}. While radial flow fluctuations are perhaps not the best observable for a precision measurement of the QGP viscosity, they provide a valuable consistency check for the hydrodynamic approach and can help to constrain the spectrum of initial-state fluctuations.

\section{ELLIPTIC AND OTHER ANISOTROPIC FLOW COEFFICIENTS}
\label{vn}

\subsection{Elliptic flow systematics}
\label{sec6.1}

Experimentally, the most direct evidence of hydrodynamic behavior comes from the observation of anisotropic flow~\cite{Ollitrault:1992bk}. The largest of the anisotropic flow coefficients is $v_2$, the elliptic flow. Like all $v_n$, it depends on $\eta/s$: Larger shear viscosity quickly reduces them. Therefore, the large elliptic flow observed at RHIC energies provides compelling evidence for strongly interacting matter that behaves like an almost perfect liquid~\cite{Arsene:2004fa,Gyulassy:2004zy}.

The viscous effects reducing the magnitude of the elliptic flow depend on the size of the system \cite{Song:2008si} which again depends on the collision centrality. The centrality dependence of the elliptic flow thus is an observable sensitive to the magnitude of $\eta/s$. In {\bf Figure~\ref{F2}} (Sec.~\ref{sec4.2}) we showed that the centrality dependence of $v_2/\varepsilon_2$ is nicely described by viscous hydrodynamic calculations. However the magnitude of $\eta/s$ used in these calculations should be considered as an average over the temperature history of the expanding fireball since we know from other fluids that $\eta/s$ depends on temperature. In addition, we also know that part of the elliptic flow originates from the hadronic phase. Therefore, knowledge of both the temperature dependence {\it and} the relative contributions from the partonic and hadronic phases is required to quantify $(\eta/s)_\mathrm{QGP}$ of the partonic fluid. 

At RHIC energies, the dissipative hadronic phase significantly affects the finally observed elliptic flow, complicating the accurate determination of $(\eta/s)_\mathrm{QGP}$. In Pb+Pb collisions at the LHC the higher collision energies produce a system that is hotter and has a longer-lived partonic phase. Consequently the hadronic contribution to the elliptic flow decreases, and this reduces the uncertainty on the determination of {$(\eta/s)_\mathrm{QGP}$.} Measurements of the energy dependence of elliptic flow at both RHIC and at the LHC allow to vary in a systematic manner the contribution from both phases and probe the temperature dependence of $\eta/s$ \cite{Niemi:2011ix}. 

\begin{figure}[thb]
\begin{minipage}{1.25\textwidth}
\includegraphics[width=0.51\textwidth]{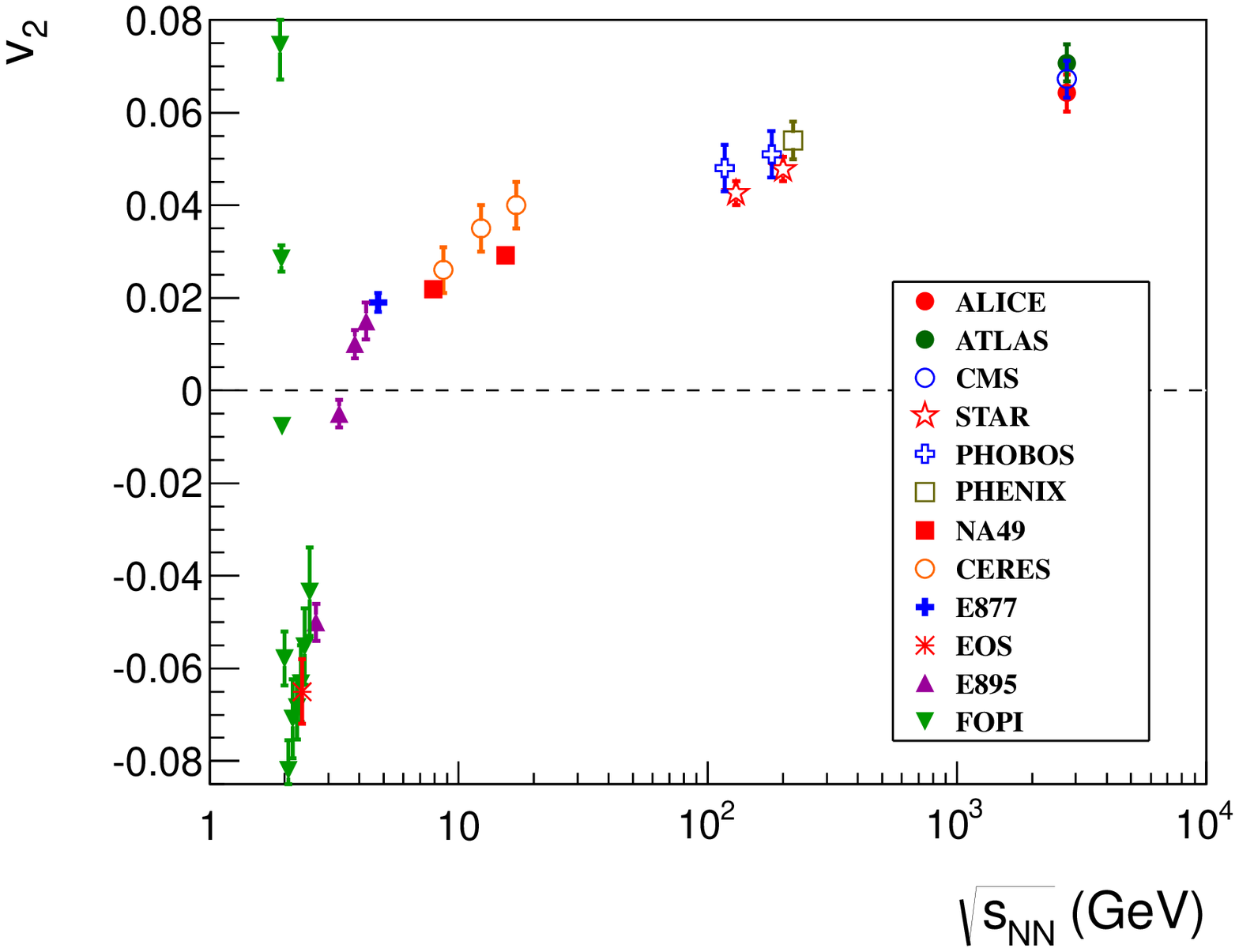}
\includegraphics[width=0.49\textwidth]{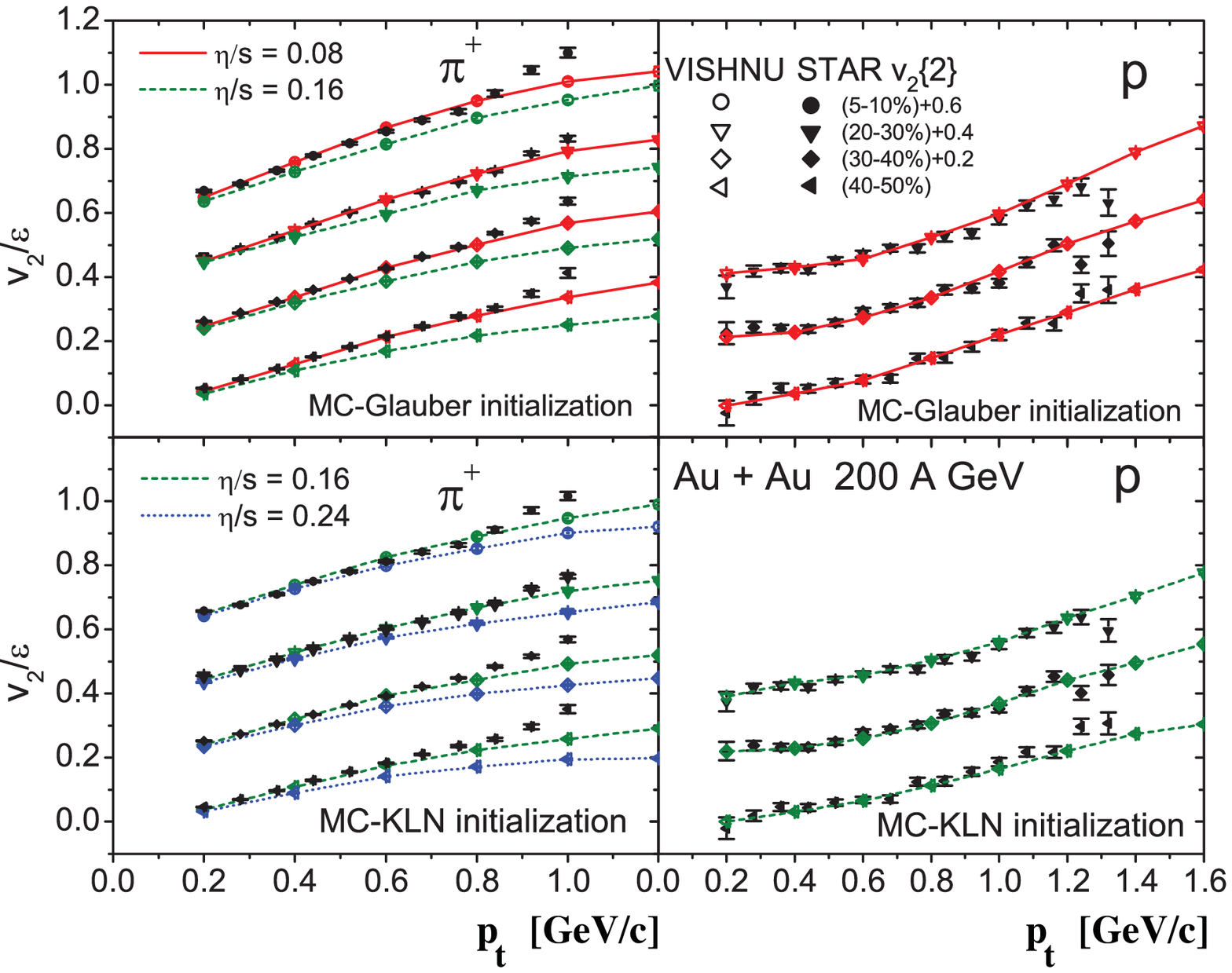}
    \caption{
    (a) Integrated elliptic flow at 2.76~TeV \cite{Aamodt:2010pa} 
    in the 20--30\% centrality class compared with results from lower energies taken at similar 
    centralities. 
    (b) The $v_2(\pt)$ for pions and protons measured by STAR compared to hydrodynamic 
    calculations with different eccentricities and $\eta/s$ \cite{Song:2011hk}.
    }
    \label{F8}
    \end{minipage}
\end{figure}

{\bf Figure~\ref{F8}a} shows the measured integrated elliptic flow at the LHC in one centrality bin, compared to results from lower energies. It shows that there is a continuous increase in the elliptic flow from {\sc RHIC} to {\sc LHC} energies.  In comparison to the elliptic flow measurements in Au+Au collisions at $\snn{\,=\,}200$\,GeV, $v_2$ increases by about 30\% at $\snn{\,=\,}2.76$\,TeV. Hydrodynamic models \cite{Niemi:2008ta,Kestin:2008bh,Luzum:2009sb} and hybrid models \cite{Hirano:2005xf,Hirano:2010jg} that successfully describe flow at RHIC indeed predicted an increase of $\sim$10--30\% in $v_2$ at the LHC. 

At low $p_T$, not only the $v_2$ of charged particles but also that of identified particles at RHIC and LHC is well described by viscous hydrodynamics. The blue shift of the transverse momenta which depends on the particle mass also generates the characteristic mass splitting observed in a plot of $v_2$ versus $p_T$ for particles of different mass \cite{Arsene:2004fa,Yin:2012sk}. {\bf Figure~\ref{F8}b} shows the pion and proton elliptic flow measured by STAR compared to {\tt VISHNU} model calculations \cite{Song:2011hk}. In the initial viscous hydrodynamic stage $\eta/s$ is taken to be temperature independent. The $\eta/s$ values required to describe the pion and proton $p_T$-differential elliptic flow data are found to be one or two times the {\sc KSS} bound \cite{Policastro:2001yc} for Glauber or KLN eccentricities, respectively, in agreement with the values required to describe the $p_T$-integrated total charged particle $v_2$.

\subsection{Triangular flow and higher flow harmonics}
\label{sec6.2}

The produced system responds as a fluid to the pressure gradients and converts the coordinate space distributions, characterized by $\varepsilon_n$, to long range momentum space correlations between the produced particles. In the last years it was realized that novel long range correlations first observed at RHIC, known as the ``ridge" and ``Mach cone", are, in fact, manifestations of anisotropic flow \cite{Sorensen:2010zq,Luzum:2010sp,ALICE:2011ab}.

Because the created matter distribution in a collision is inhomogeneous, $n$ covers a large range ({\bf Figure~\ref{F2}}). However, the shear viscosity reduces differences between the expansion velocities and therefore dampens the anisotropic flow coefficients $v_n$, and more strongly so for larger $n$. As a consequence, the magnitude and transverse momentum dependence of the $v_n$ coefficients provide a large set of observables \cite{ALICE:2011ab,ATLAS:2012at,CMS,Adare:2011tg,Pandit:2012mq} to check the hydrodynamical paradigm and are, within this description, very sensitive to the magnitude of $\eta/s$~\cite{Alver:2010dn,Schenke:2011zz}.

\begin{figure}[thb]
\begin{minipage}{1.25\textwidth}
    \includegraphics[width=0.48\textwidth]{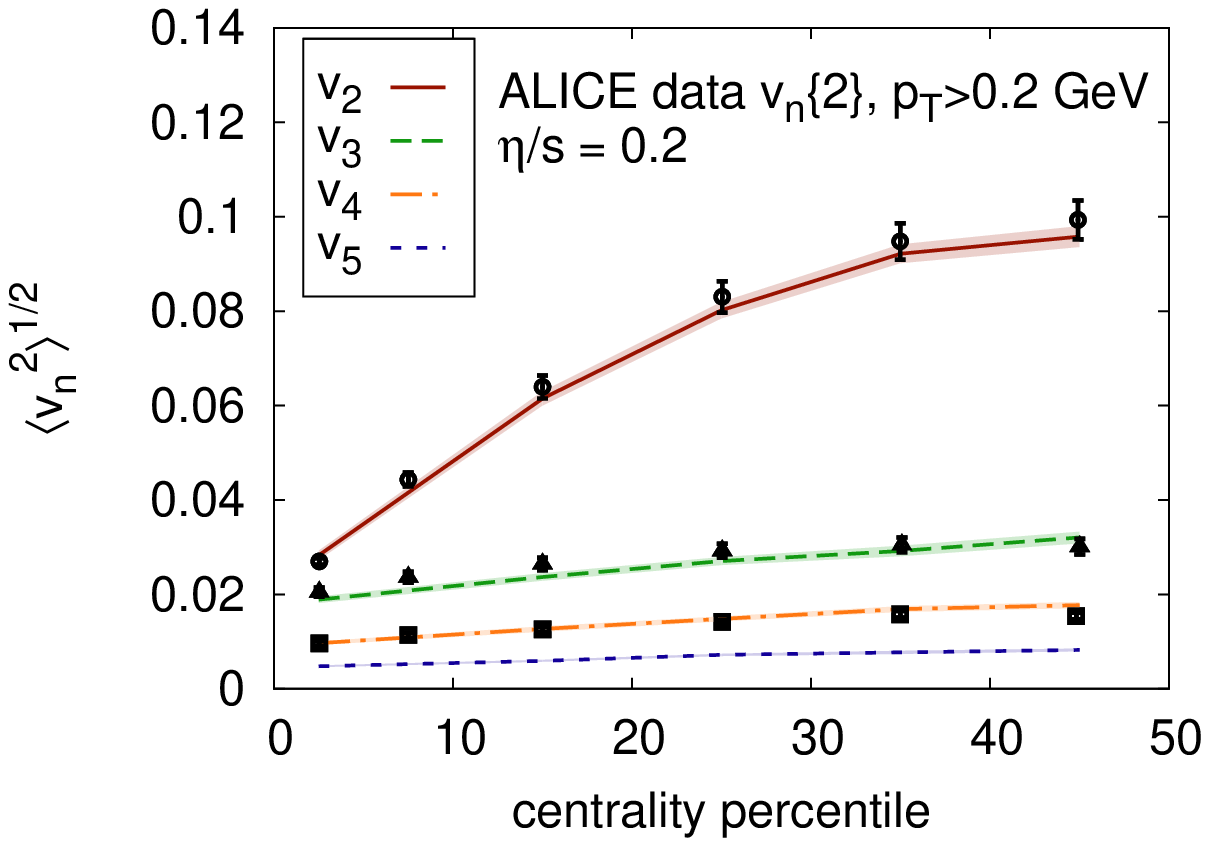}
    \includegraphics[width=0.52\textwidth]{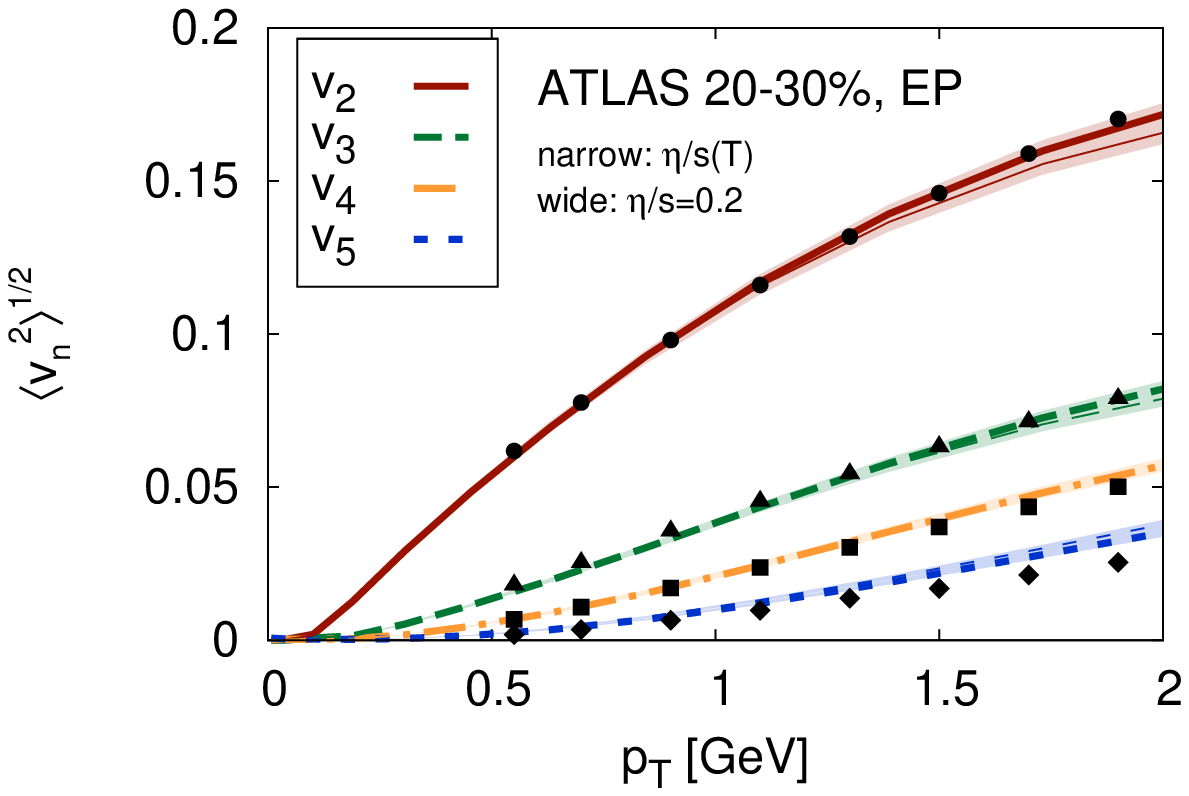}
    \caption{
    (a) The centrality dependence of $v_n\{2\}$ from 2.76\,$A$\,TeV Pb+Pb collisions measured by ALICE \cite{ALICE:2011ab} compared to viscous hydrodynamic model calculations \cite{Gale:2012rq}.
    (b) Comparison of $v_n(\pt)$ for the same collision system at $20{-}30\%$ centrality from ATLAS \cite{ATLAS:2012at} with hydrodynamical calculations, 
    using both a constant average and a temperature dependent $\eta/s$ \cite{Gale:2012rq}.
    }
    \label{F9}
    \end{minipage}
\end{figure}

In {\bf Figure~\ref{F9}a} the measured $v_2$, $v_3$ and $v_4$ are shown as functions of centrality \cite{Gale:2012rq}. These form the {\em flow power spectrum} created by hydrodynamics in response to the {\em initial fluctuation power spectrum} shown in {\bf Figure~\ref{F2}}. $p_T$-integrated charged hadron flow coefficients for $n{\,=\,}1$ and $n{\,>\,}4$ are not yet available experimentally but have been predicted theoretically \cite{Gale:2012rq} and are expected to complement the Little Bang power spectrum in the near future. Clearly $v_3$ and $v_4$ have a rather weak centrality dependence while $v_2$ changes rapidly. This mirrors the spectrum of $\varepsilon_n$ in {\bf Figure~\ref{F2}} which also shows the strongest centrality dependence for $n{\,=\,2}$. Within hydrodynamics, the initial $\varepsilon_n$ power spectrum together with the shear viscosity $(\eta/s)(T)$ completely determine the centrality and $p_T$ dependences of the anisotropic flow coefficients $v_n$, as well as their event-by-event fluctuations. The IP-Glasma initial conditions \cite{Schenke:2012wb}, together with an average value for $\eta/s$ of 0.2 for Pb+Pb collisions at the LHC and a somewhat smaller value of 0.12 for Au+Au collisions at top RHIC energies, provide a good description of all presently available data for charged hadron $v_n$ and $v_n(p_T)$ \cite{Gale:2012rq}. Examples are shown in {\bf Figure~\ref{F9}}. As discussed in \cite{Gale:2012rq}, the measurements at both collision energies are also compatible with a temperature dependent specific shear viscosity $(\eta/s)(T)$ \cite{Niemi:2011ix} that has a minimum value of $\frac{1}{4\pi}{\,=\,}0.08$ at $T_c$ and rises moderately above and more steeply below $T_c$ ({\it c.f.} the two curves shown in comparison with $v_2(p_T)$ data in {\bf Figure~\ref{F9}b}). On the other hand, initial fluctuation power spectra obtained from the MC-Glauber and MC-KLN models (also shown in {\bf Figure~\ref{F2}}) can not reproduce the measured flow power spectrum (C.\ Shen and Z.\ Qiu, private communication, and related discussion in \cite{Jia:2012ve}). Although a good simultaneous description of charged hadron $v_2$ and $v_3$ can be obtained with $\eta/s=0.08$ and MC-Glauber initial conditions \cite{Qiu:2011hf} (but not with MC-KLN initial conditions), both models fail when tested on higher flow harmonics and the widths of their event-by-event fluctuation distributions.
 
Fluctuations also induce a non-zero directed flow $v_1$ at midrapidity \cite{Teaney:2010vd,Gardim:2011qn}. It is strongly constrained by global momentum conservation \cite{Gardim:2011qn} (which forces the $p_T$-weighted directed flow to vanish at midrapidity) and has been measured both at RHIC \cite{Pandit:2012hp} and LHC \cite{Selyuzhenkov:2011zj}. Its effect is clearly visible in {\bf Figure~\ref{F1}} as a dipole shift of the matter density. Once still existing differences between the measurements have been clarified, this observable will form the bottom end of the flow power spectrum of the Little Bang.
 
\subsection{Flow angle correlations}
\label{sec6.3}

The anisotropic flow coefficients $v_2$ and $v_3$ are to a good approximation linearly proportional to the $\varepsilon_2$ and $\varepsilon_3$, respectively~\cite{Qiu:2011iv}. However, due to the nonlinearity of the hydrodynamic evolution equations, the hydrodynamic response to the initial-state fluctuations spectrum is characterized by mode-mixing between different order flow harmonics which becomes large in mid-central and peripheral collisions where the large geometric ellipticity $\varepsilon_2$ drives a large elliptic flow $v_2$ which mixes with the other flow harmonics \cite{Qiu:2011iv,Gardim:2011xv,Teaney:2012ke}. Anisotropic flow coefficients such as $v_4$ and $v_5$ thus depend on the magnitude of $v_2$ and $v_3$, respectively, and for mid-central collisions the flow angles $\Psi_4$ and $\Psi_5$ are uncorrelated with the participant plane angles $\Phi_4$ and $\Phi_5$ associated with $\varepsilon_4$ and $\varepsilon_5$ \cite{Qiu:2011iv}. This characteristic change between the initial coordinate space $\varepsilon_n$ and final momentum space $v_n$ power spectra and their associated angles is a strong test of the hydrodynamic paradigm and provides additional constraints on the specific shear viscosity and initial density fluctuation spectrum~\cite{Qiu:2012uy}.
 
Correlations between the flow planes $\Psi_n$ can be measured either by using combinations of two particle correlations to estimate each $\Psi_n$, or by using multi-particle cumulants.
%
\begin{figure}
\begin{minipage}{1.25\textwidth}
\includegraphics[width=0.55\textwidth]{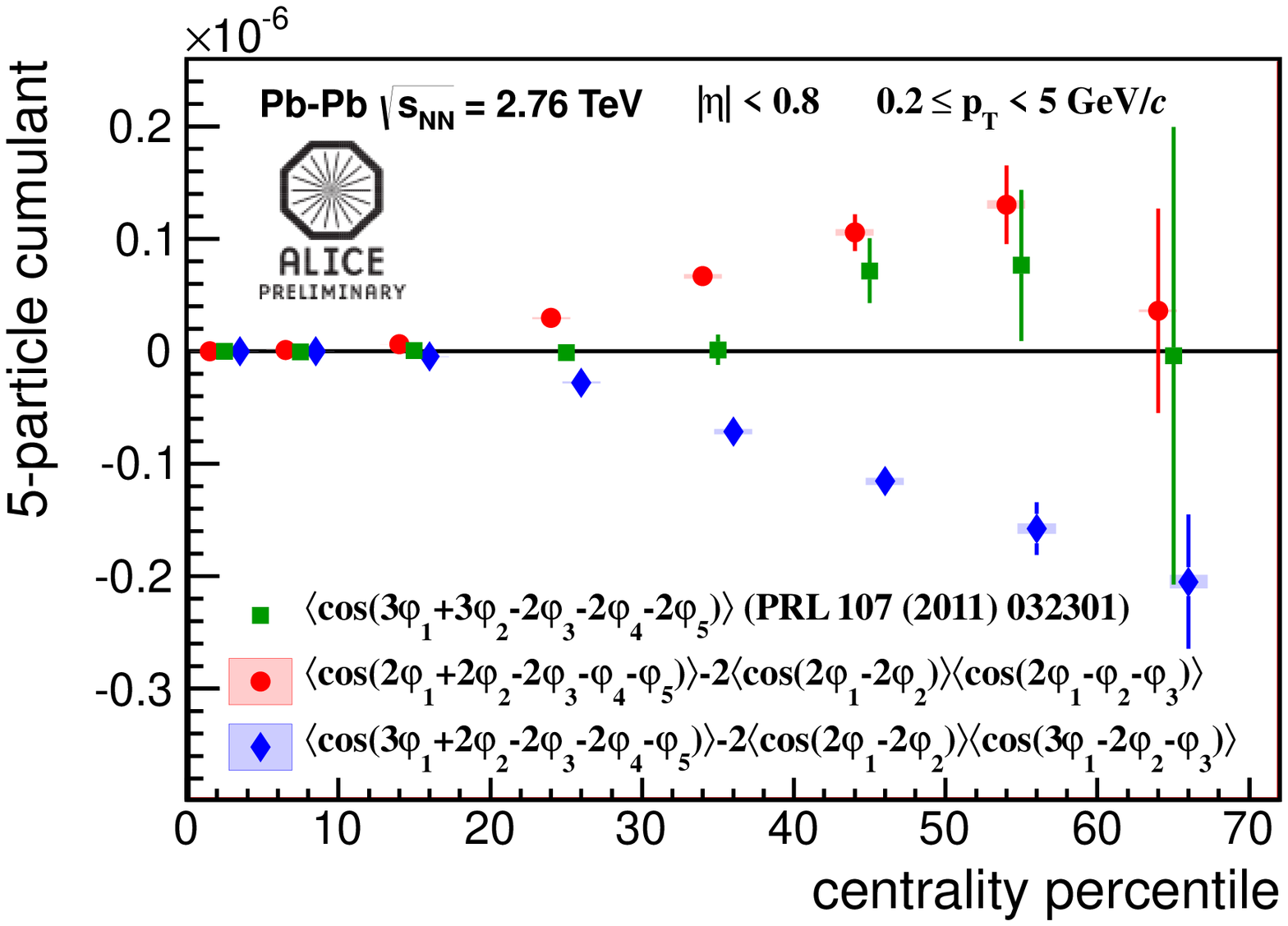}
\includegraphics[width=0.45\textwidth]{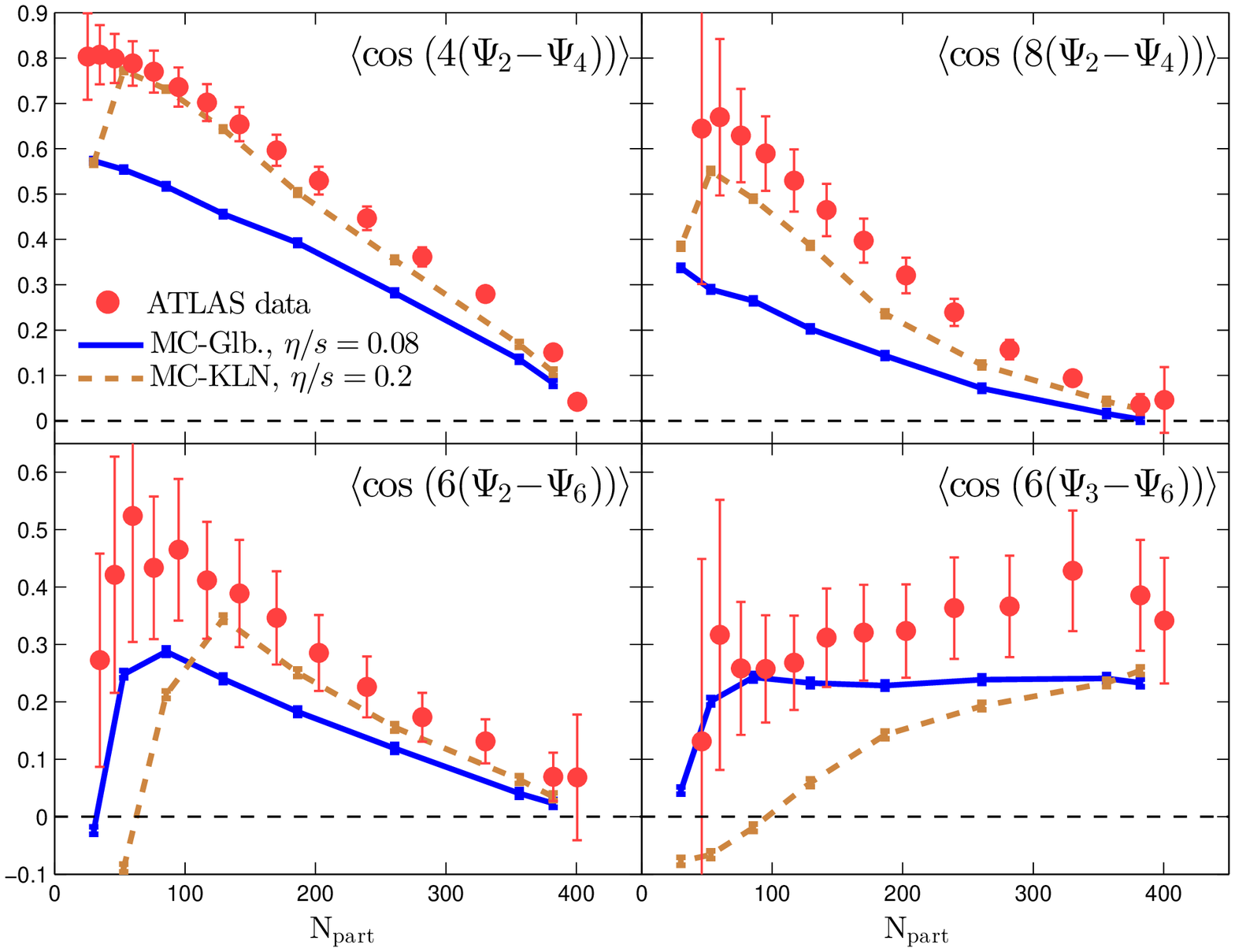}
\caption{
(a) The centrality dependence of the 5-particle cumulants from ALICE \cite{ALICE:2011ab}. The cumulant shown by green markers is sensitive to $v_3^2v_2^3 \cos [6(\Psi_3{-}\Psi_2)]$, the red markers to $-v_2^3v_1^2 \cos [2(\Psi_2{-}\Psi_1)]$, and the blue markers to $-v_3 v_2^3 v_1 \cos(3\Psi_3{-}2\Psi_2{-}\Psi_1)$. 
(b) Correlations between different $\Psi_n$ from ATLAS \cite{Jia:2012sa} and from viscous hydrodynamics \cite{Qiu:2012uy}. 
Both data sets are from Pb+Pb collisions at the LHC. 
}
\label{F10} 
\end{minipage}
\end{figure}
%
{\bf Figure~\ref{F10}a} shows the ALICE measurements \cite{ALICE:2011ab} of 5-particle cumulants that are sensitive to the correlation among different order flow planes. The correlator $\dla\cos[3(\varphi_1{+}\varphi_2) - 2(\varphi_3{+}\varphi_4{+}\varphi_5)]\dra$ (where $\dla\dots\dra$ denotes the double average over the particles 1--5 within an event and over all events) is sensitive to $\langle v_3^2v_2^3 \cos[6(\Psi_3{-}\Psi_2)]\rangle$ (where $\langle\dots\rangle$ denotes an average over events); the data show that $\Psi_3$ and $\Psi_2$ are uncorrelated for central to mid-central collisions. However, the three-plane correlation between the angles $\Psi_1$, $\Psi_2$ and $\Psi_3$, obtained by measuring the five-particle cumulant shown by the blue markers, is already significant for mid-central collisions, in qualitative agreement with expectations from (linearized) hydrodynamic response \cite{Teaney:2010vd}.

In general, however, linearized hydrodynamic response is not sufficient. In {\bf Figure~\ref{F10}b} we show several correlation functions between flow angles corresponding to different harmonics as measured by ATLAS~\cite{Jia:2012sa}, plotted versus centrality, with central collisions (large $N_{\rm part}$) on the the right and peripheral collisions (small $N_{\rm part}$) on the left. The comparison to viscous hydrodynamic calculations \cite{Qiu:2012uy} with initial energy density profiles from the MC-Glauber and MC-KLN models (solid and dashed lines) shows good qualitative overall agreement; the corresponding correlations between the eccentricity planes $\Phi_n$ in the initial state, on the other hand, behave quite differently -- in magnitude, in their qualitative centrality dependence, and even in sign \cite{Qiu:2012uy}. The final-state flow angle correlations thus cannot be understood, even qualitatively, without taking into account the nonlinear hydrodynamic response of the QGP fluid to the fluctuating initial states \cite{Qiu:2012uy,Teaney:2012ke}. The strength of these nonlinear effects and, in some cases, even the shape of the centrality dependence of these flow angle correlations were shown to be sensitive both to the details of the initial-state fluctuation spectrum and to the specific shear viscosity $\eta/s$ \cite{Qiu:2012uy}.

\subsection{Eccentricity and flow fluctuations}
\label{sec6.4}

As explained in Sec.~\ref{sec3.1}, experimentally the anisotropic flow coefficients are estimated from measured angular correlations between emitted particles. Different correlation functions probe different moments of the $v_n$ distributions, but none of them measures directly the mean $\bar{v}_n$. As seen in Eq.~(\ref{eq3.5}), the mean and variance of $v_n$ can be approximately obtained from the two- and four-particle cumulants (after correcting the former for non-flow effects \cite{Ollitrault:2009ie}) as $\bar{v}_n \approx \sqrt{(v_n^2\{2\}{+}v_n^2\{4\})/2}$ and $\sigma_{v_n} \approx \sqrt{(v_n^2\{2\}{-}v_n^2\{4\})/2}$.\footnote{Here we ignored flow angle fluctuations by setting $\bar{v}_n{\,=\,}\langle v_n\rangle$.} 
%
\begin{figure}[ht]
\begin{minipage}{1.25\textwidth}
\includegraphics[width=0.6\textwidth]{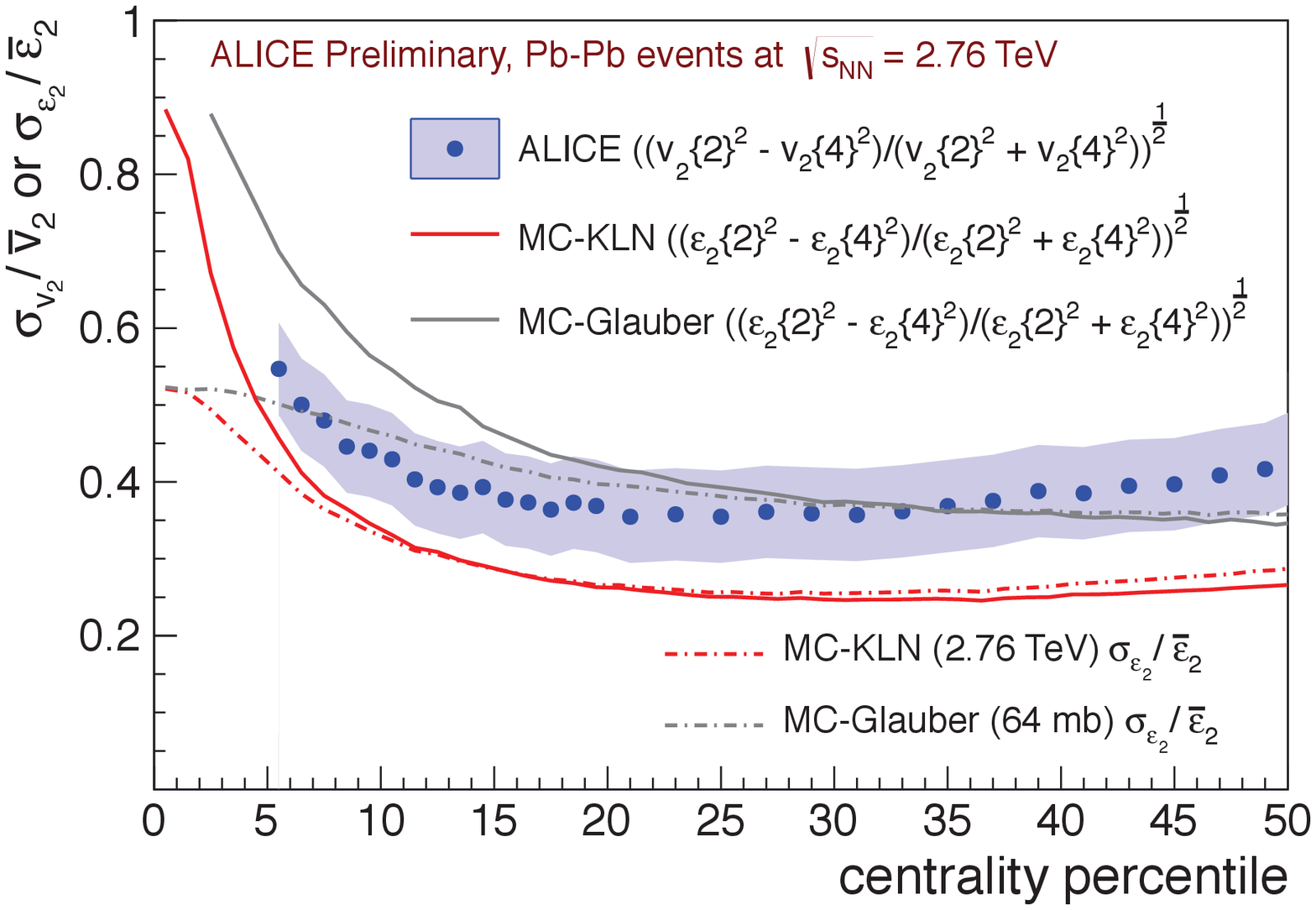}
\includegraphics[width=0.4\textwidth]{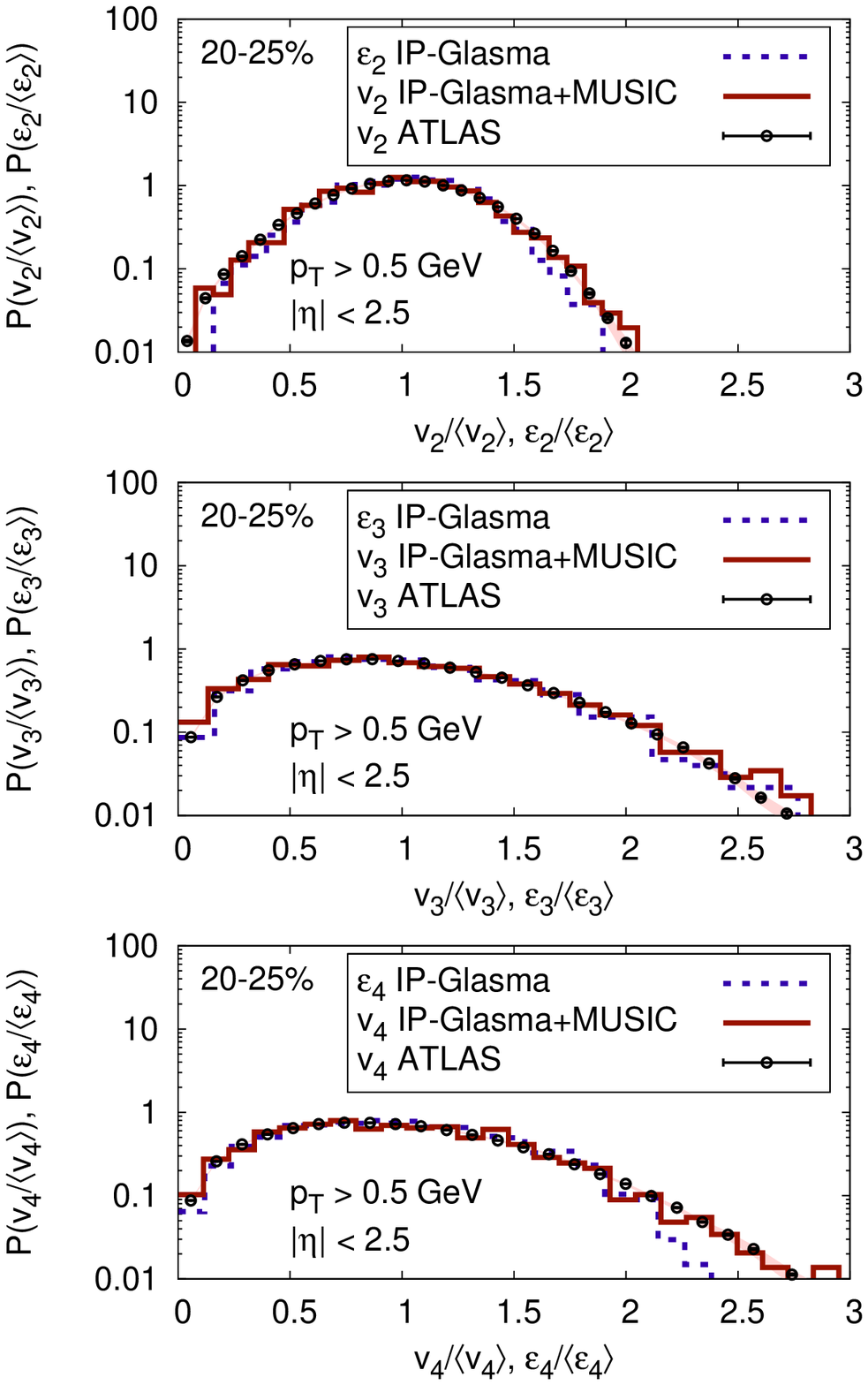}
\caption{
(a) Relative event-by-event elliptic flow fluctuations versus collision centrality measured by ALICE~\cite{Collaboration:2011yba}. 
(b) Scaled distributions of $v_{2,3,4}$ (from top to bottom) from viscous hydrodynamics with IP-Glasma initial conditions \cite{Gale:2012rq} compared with experimental data from ATLAS~\cite{Jia:2012ve} and with the scaled distributions of the corresponding initial eccentricities $\varepsilon_{2,3,4}$. Nonlinear hydrodynamic evolution causes slightly larger variances for the $v_n$ distributions compared to those of $\varepsilon_n$. The data in both panels are from Pb+Pb collisions at the LHC.
}
\label{F11} 
\end{minipage}
\end{figure}
%
The blue markers in {\bf Figure~\ref{F11}a} show the resulting experimental estimate for the normalized elliptic flow variance $\sigma_{v_2}/\bar{v}_2$. It is found to be large (${\approx\,}40\%$), hence the shown $\bar{v}_2$ and $\sigma_{v_2}$ are only rough estimates. Since, for not too large impact parameters \cite{Qiu:2011iv}, the magnitude of the elliptic flow is proportional to the eccentricity $\varepsilon_2$ of the initial nuclear overlap region, we expect (and the top panel in {\bf Figure\ref{F11}b} confirms) that the normalized variance of $v_2$ should be close to that of $\varepsilon_2$, shown by the lines in {\bf Figure~\ref{F11}} for MC-Glauber \cite{Miller:2007ri} and MC-KLN \cite{Kharzeev:2001yq} initial conditions. The difference between the dash-dotted and solid lines, showing the exact ratio $\sigma_{\varepsilon_2}/\bar{\varepsilon}_2$ from the respective model and its small-variance approximation, indicates the quality of the approximation used in the experimental extraction of $\sigma_{v_2}/\bar{v}_2$. The rise of the experimental estimate for $\sigma_{v_2}/\bar{v}_2$ in central collisions is seen to be mostly due to this approximation. Although the centrality dependences of $\sigma_{\varepsilon_2}/\bar{\varepsilon}_2$ from the models and of $\sigma_{v_2}/\bar{v}_2$ from experiment show clear similarities, neither model is able to match the data over the full centrality range. 

For a more detailed comparison the full event-by-event distribution of $v_n$ can be reconstructed experimentally, by removing as much as possible non-flow correlations and then unfolding the measured correlations with a response function that accounts for the statistical smearing due to the finite number of measured particles per event \cite{Jia:2012ve}. {\bf Figure~\ref{F11}b} compares the measured event-by-event distributions of $v_2$, $v_3$ and $v_4$ measured by ATLAS~\cite{Jia:2012ve}, the distribution of the $\varepsilon_n$ from the IP-Glasma model \cite{Gale:2012rq}, and the final $v_n$ distributions after the viscous hydrodynamical evolution. The theoretical $v_n$ distributions after hydrodynamic evolution are seen to be slightly wider than the initial $\varepsilon_n$ distributions, indicating non-linear hydrodynamic effects on the fluctuation spectrum. These appear to be somewhat stronger in peripheral than in central collisions \cite{Gale:2012rq}, which may explain the stronger increase of $\sigma_{v_2}/\bar{v}_2$ compared to $\sigma_{\varepsilon_2}/\bar{\varepsilon}_2$ seen in {\bf Figure~\ref{F11}a} for large impact parameters. In {\bf Figure~\ref{F11}b} non-linear mode-mixing effects on the shape of the event-by-event distribution are particularly prominent for $v_4$; only after accounting for the non-linear hydrodynamic evolution, the measured $v_n$ distributions can be fully reproduced.

The ability of the IP-Glasma initial-state fluctuation model, which is deeply rooted in QCD, combined with viscous hydrodynamic evolution to simultaneously correctly describe the mean $v_{2,3,4}$ coefficients at both top RHIC and LHC energies, their centrality and $p_T$ dependence, as well as the detailed shape of their event-by-event distributions, is impressive. The recent measurements of these observables and their successful theoretical explanation has established the hydrodynamic paradigm for ultra-relativistic heavy-ion collisions beyond reasonable doubt. Future measurements of the centrality and beam energy dependence of the complete $v_n$ power spectrum and of the correlations between the flow angles $\Psi_n$, together with more complete theoretical modeling of the early-pre-equilibrium and late freeze-out stages, will allow for a precision determination of the transport properties of the QGP and the final completion of the {\it Little Bang Standard Model}.

\section{SYNOPSIS AND FUTURE PERSPECTIVE}
\label{synopsis}

Construction work on the {\em Little Bang Standard Model} is nearing completion. Heavy-ion collisions generate many different classes of Little Bangs -- the initial conditions depend on collision system, collision energy and collision centrality. Each class is characterized by its own spectrum of initial-state density fluctuations which (at LHC and top RHIC energy) appear to be calculable directly from QCD, using gluon saturation ideas. This so-called Glasma model also provides a description of the early pre-equilibrium evolution of the Little Bang, although the absence of color field quantum fluctuations in the present implementations, and the resulting lack of thermalization, are weaknesses that need to be fixed -- corresponding work is going on while this is being written. After a very short time of around $0.2{-}0.4$\,fm/$c$ (for top RHIC and LHC energies), the Glasma pre-equilibrium dynamics can be matched to viscous fluid dynamics of an approximately thermalized QGP, which then describes the main part of the fireball evolution (the next $5{-}10$\,fm/$c$, depending on collision energy and geometry) until hadronization. After hadronization, the macroscopic hydrodynamic evolution should be matched to a microscopic kinetic description based on hadronic degrees of freedom which then describes the rest of the Little Bang's life until final decoupling. At LHC energies, replacing this microscopic kinetic approach by a continued application of fluid dynamics, with appropriately adjusted hadronic chemical potentials in the equation of state that describe the breaking of hadronic chemical equilibrium below $T_c$, and with appropriately chosen values for the specific shear and bulk viscosities $\eta/s$ and $\zeta/s$ in the hadronic phase, is for many observables a reasonably good approximation, but will in the end not suffice for precision work. 

The fact that the initial Little Bang density profiles fluctuate from event to event is of crucial importance. Fluctuations in the magnitude and orientation of the initial-state eccentricity coefficients lead to fluctuations in the final harmonic flow coefficients and in the directions of those flows whose variances affect measured quantities in specific ways that require, for a faithful theoretical description, event-by-event simulation of the hydrodynamic evolution. The correlations between the final flow planes differ from those between the initial eccentricity planes, sometimes dramatically, reflecting the non-linear hydrodynamic evolution and mode-mixing, in particular with the large geometrically driven elliptic flow in peripheral collisions. The recent measurement of these correlations and their successful explanation with full non-linear hydrodynamics, but not with linear response theory, constitutes an {\em experimentum crucis} in support of the hydrodynamic paradigm. The community is looking forward to fully exploiting the spectrum of final anisotropic flow fluctuations and their angular correlations for pinning down the initial fluctuation spectrum and QGP transport properties.

The present, still imperfect version of the Little Bang Standard Model is able to describe qualitatively, and in most cases even quantitatively, almost all observed features of soft hadron production in heavy-ion collisions at RHIC and LHC. The fact that collective flow anisotropies are uniquely sensitive the shear viscosity of the fireball matter has permitted the experimental determination of $(\eta/s)_\mathrm{QGP}\approx 0.2 = 2.5\times\frac{1}{4\pi}$ at LHC temperatures (perhaps a bit smaller at RHIC temperatures), with less than 20\% statistical and of order 50\% systematic uncertainty related to the initial-state fluctuation and pre-equilibrium evolution models. Rapid recent progress and ongoing work promise to reduce, within a year or two, the theoretical systematic error to a level competitive with the experimental statistical and systematic errors. Together with additional high-precision experiments, further exploration of the systematic variation of flow observables with collision energy, impact parameter and system size should permit the determination of $(\eta/s)_\mathrm{QGP}$ with 5\% relative precision, accurate enough to open the window for an experimental investigation of the (by an order of magnitude) smaller effects from bulk viscosity and the related shear and bulk viscous relaxation times.   

\bigskip

\noindent
{\bf Acknowledgements:} We thank Bjoern Schenke, Zhi Qiu and Chun Shen for help in preparing Figure~\ref{F2} and You Zhou for the preparation of Figure \ref{F4}. Constructive comments by P.\ Huovinen and J.\ Jia are gratefully acknowledged. The work of UH was supported by the US Department of Energy under Grants No. {\rm DE-SC0004286} and (within the framework of the JET Collaboration) {\rm DE-SC0004104}. The work of RS was supported by the Stichting voor Fundamenteel Onderzoek der Materie (FOM) and the Nederlandse Organisatie voor Wetenschappelijk Onderzoek (NWO).




\end{document}